\newcommand{\citey}[1]{\citeauthor{#1} \citeyear{#1}}

\documentclass[twocolumn, tighten]{aastex631} 
\usepackage{booktabs}
\usepackage{multirow}

\usepackage{amsmath}
\usepackage{enumitem}

\usepackage{bm}
\usepackage{verbatim}

\usepackage{tabularx} 
\usepackage[utf8]{inputenc}
\usepackage{textcomp}

\journalinfo{Published in ApJ, 1005, 200 (2026)}

\begin{document}

\title{The Kinematically Hot, Extremely Metal-Poor C-19 Stellar Stream in DESI DR2}
\shorttitle{C-19 in DESI}
\shortauthors{N. Mohammed et al.}

\correspondingauthor{Nasser Mohammed}

\author[0009-0008-1224-0382]{Nasser Mohammed}
\affiliation{David A. Dunlap Department of Astronomy \& Astrophysics, University of Toronto, 50 St. George Street, Toronto, ON, M5S 3H4, Canada}
\affiliation{Dunlap Institute for Astronomy \& Astrophysics, 50 St. George Street, Toronto, ON, M5S 3H4, Canada}\email{nasser.mohammed@astro.utoronto.ca}

\author[0009-0006-2841-8101]{Joseph~Y.~Tang}
\affiliation{David A. Dunlap Department of Astronomy \& Astrophysics, University of Toronto, 50 St. George Street, Toronto, ON, M5S 3H4, Canada}
\affiliation{Department of Astronomy, Columbia University, New York, NY, 10027, USA}

\author[0000-0002-9110-6163]{Ting S. Li}
\affiliation{David A. Dunlap Department of Astronomy \& Astrophysics, University of Toronto, 50 St. George Street, Toronto, ON, M5S 3H4, Canada}
\affiliation{Dunlap Institute for Astronomy \& Astrophysics, 50 St. George Street, Toronto, ON, M5S 3H4, Canada}
\email{ting.li@astro.utoronto.ca}

\author[0000-0003-2644-135X]{Sergey E. Koposov}
\affiliation{Institute for Astronomy, University of Edinburgh, Royal Observatory, Blackford Hill, Edinburgh EH9 3HJ, UK}
\affiliation{Institute of Astronomy, University of Cambridge, Madingley Road, Cambridge CB3 0HA, UK}

\author[0000-0002-7667-0081]{Raymond G. Carlberg}
\affiliation{David A. Dunlap Department of Astronomy \& Astrophysics, University of Toronto, 50 St. George Street, Toronto, ON, M5S 3H4, Canada}

\author[0009-0006-5612-7336]{Emma Jarvis}
\affiliation{David A. Dunlap Department of Astronomy \& Astrophysics, University of Toronto, 50 St. George Street, Toronto, ON, M5S 3H4, Canada}
\affiliation{Dunlap Institute for Astronomy \& Astrophysics, 50 St. George Street, Toronto, ON, M5S 3H4, Canada}

\author[0009-0005-5355-5899]{Andrew~P.~Li}
\affiliation{David A. Dunlap Department of Astronomy \& Astrophysics, University of Toronto, 50 St. George Street, Toronto, ON, M5S 3H4, Canada}
\affiliation{Physics \& Astronomy Department, University of Victoria, 3800 Finnerty Road, Victoria, BC, V8P 5C2, Canada}

\author[0000-0002-7393-3595]{Nathan Sandford}
\affiliation{David A. Dunlap Department of Astronomy \& Astrophysics, University of Toronto, 50 St. George Street, Toronto, ON, M5S 3H4, Canada}

\author[0000-0003-0105-9576]{Gustavo~E.~Medina}
\affiliation{David A. Dunlap Department of Astronomy \& Astrophysics, University of Toronto, 50 St. George Street, Toronto, ON, M5S 3H4, Canada}
\affiliation{Dunlap Institute for Astronomy \& Astrophysics, 50 St. George Street, Toronto, ON, M5S 3H4, Canada}

\author[0000-0002-5762-7571]{Wenting Wang}
\affiliation{Department of Astronomy, School of Physics and Astronomy, and Key Laboratory for Particle Astrophysics and Cosmology (MOE)/Shanghai Key Laboratory for Particle Physics and Cosmology, Shanghai Jiao Tong University, Shanghai 200240, People's Republic of China}

\author[0000-0002-6257-2341]{Monica Valluri}
\affiliation{Department of Astronomy and Astrophysics, University of Michigan, Ann Arbor, MI, USA}

\author[0000-0001-5805-5766]{Alexander H.~Riley}
\affiliation{Lund Observatory, Division of Astrophysics, Department of Physics, Lund University, SE-221 00 Lund, Sweden}

\author[0000-0002-0740-1507]{Leandro {Beraldo e Silva}}
\affiliation{Observatório Nacional, Rio de Janeiro - RJ, 20921-400, Brasil}

\author[0000-0002-5758-150X]{Joan Najita}
\affiliation{NSF’s NOIRLab, 950 N. Cherry Avenue, Tucson, AZ 85719, USA}

\author[0000-0002-2527-8899]{Mika Lambert}
\affiliation{Department of Astronomy \& Astrophysics, University of California, Santa Cruz, 1156 High Street, Santa Cruz, CA 95064, USA}

\author[0000-0002-6469-8263]{Songting Li}
\affiliation{Department of Astronomy, School of Physics and Astronomy, and Key Laboratory for Particle Astrophysics and Cosmology (MOE)/Shanghai Key Laboratory for Particle Physics and Cosmology, Shanghai Jiao Tong University, Shanghai 200240, People's Republic of China}
\affiliation{Tsung-Dao Lee Institute, Shanghai Jiao Tong University, Shanghai, 201210, China}
\affiliation{State Key Laboratory of Dark Matter Physics, School of Physics and Astronomy, Shanghai Jiao Tong University, Shanghai 200240, China}

\author{J.~Aguilar}
\affiliation{Lawrence Berkeley National Laboratory, 1 Cyclotron Road, Berkeley, CA 94720, USA}

\author[0000-0001-6098-7247]{S.~Ahlen}
\affiliation{Department of Physics, Boston University, 590 Commonwealth Avenue, Boston, MA 02215 USA}

\author[0000-0001-9712-0006]{D.~Bianchi}
\affiliation{Dipartimento di Fisica ``Aldo Pontremoli'', Universit\`a degli Studi di Milano, Via Celoria 16, I-20133 Milano, Italy}
\affiliation{INAF-Osservatorio Astronomico di Brera, Via Brera 28, 20122 Milano, Italy}

\author{D.~Brooks}
\affiliation{Department of Physics \& Astronomy, University College London, Gower Street, London, WC1E 6BT, UK}

\author{T.~Claybaugh}
\affiliation{Lawrence Berkeley National Laboratory, 1 Cyclotron Road, Berkeley, CA 94720, USA}

\author[0000-0001-8274-158X]{A.~P.~Cooper}
\affiliation{Institute of Astronomy and Department of Physics, National Tsing Hua University, 101 Kuang-Fu Rd. Sec. 2, Hsinchu 30013, Taiwan}

\author[0000-0002-1769-1640]{A.~de la Macorra}
\affiliation{Instituto de F\'{\i}sica, Universidad Nacional Aut\'{o}noma de M\'{e}xico,  Circuito de la Investigaci\'{o}n Cient\'{\i}fica, Ciudad Universitaria, Cd. de M\'{e}xico  C.~P.~04510,  M\'{e}xico}

\author[0000-0002-2890-3725]{J.~E.~Forero-Romero}
\affiliation{Departamento de F\'isica, Universidad de los Andes, Cra. 1 No. 18A-10, Edificio Ip, CP 111711, Bogot\'a, Colombia}
\affiliation{Observatorio Astron\'omico, Universidad de los Andes, Cra. 1 No. 18A-10, Edificio H, CP 111711 Bogot\'a, Colombia}

\author[0000-0001-9632-0815]{E.~Gaztañaga}
\affiliation{Institut d'Estudis Espacials de Catalunya (IEEC), c/ Esteve Terradas 1, Edifici RDIT, Campus PMT-UPC, 08860 Castelldefels, Spain}
\affiliation{Institute of Cosmology and Gravitation, University of Portsmouth, Dennis Sciama Building, Portsmouth, PO1 3FX, UK}
\affiliation{Institute of Space Sciences, ICE-CSIC, Campus UAB, Carrer de Can Magrans s/n, 08913 Bellaterra, Barcelona, Spain}

\author[0000-0003-3142-233X]{S.~Gontcho A Gontcho}
\affiliation{University of Virginia, Department of Astronomy, Charlottesville, VA 22904, USA}

\author{G.~Gutierrez}
\affiliation{Fermi National Accelerator Laboratory, PO Box 500, Batavia, IL 60510, USA}

\author[0000-0003-0201-5241]{R.~Joyce}
\affiliation{NSF NOIRLab, 950 N. Cherry Ave., Tucson, AZ 85719, USA}

\author[0000-0002-0000-2394]{S.~Juneau}
\affiliation{NSF NOIRLab, 950 N. Cherry Ave., Tucson, AZ 85719, USA}

\author{R.~Kehoe}
\affiliation{Department of Physics, Southern Methodist University, 3215 Daniel Avenue, Dallas, TX 75275, USA}

\author[0000-0003-3510-7134]{T.~Kisner}
\affiliation{Lawrence Berkeley National Laboratory, 1 Cyclotron Road, Berkeley, CA 94720, USA}

\author[0000-0001-6356-7424]{A.~Kremin}
\affiliation{Lawrence Berkeley National Laboratory, 1 Cyclotron Road, Berkeley, CA 94720, USA}

\author[0000-0003-1838-8528]{M.~Landriau}
\affiliation{Lawrence Berkeley National Laboratory, 1 Cyclotron Road, Berkeley, CA 94720, USA}

\author[0000-0001-7178-8868]{L.~Le~Guillou}
\affiliation{Sorbonne Universit\'{e}, CNRS/IN2P3, Laboratoire de Physique Nucl\'{e}aire et de Hautes Energies (LPNHE), FR-75005 Paris, France}

\author[0000-0003-4962-8934]{M.~Manera}
\affiliation{Departament de F\'{i}sica, Serra H\'{u}nter, Universitat Aut\`{o}noma de Barcelona, 08193 Bellaterra (Barcelona), Spain}
\affiliation{Institut de F\'{i}sica d’Altes Energies (IFAE), The Barcelona Institute of Science and Technology, Edifici Cn, Campus UAB, 08193, Bellaterra (Barcelona), Spain}

\author[0000-0002-1125-7384]{A.~Meisner}
\affiliation{NSF NOIRLab, 950 N. Cherry Ave., Tucson, AZ 85719, USA}

\author{R.~Miquel}
\affiliation{Instituci\'{o} Catalana de Recerca i Estudis Avan\c{c}ats, Passeig de Llu\'{\i}s Companys, 23, 08010 Barcelona, Spain}
\affiliation{Institut de F\'{i}sica d’Altes Energies (IFAE), The Barcelona Institute of Science and Technology, Edifici Cn, Campus UAB, 08193, Bellaterra (Barcelona), Spain}

\author[0000-0001-9070-3102]{S.~Nadathur}
\affiliation{Institute of Cosmology and Gravitation, University of Portsmouth, Dennis Sciama Building, Portsmouth, PO1 3FX, UK}

\author[0000-0002-0644-5727]{W.~J.~Percival}
\affiliation{Department of Physics and Astronomy, University of Waterloo, 200 University Ave W, Waterloo, ON N2L 3G1, Canada}
\affiliation{Perimeter Institute for Theoretical Physics, 31 Caroline St. North, Waterloo, ON N2L 2Y5, Canada}
\affiliation{Waterloo Centre for Astrophysics, University of Waterloo, 200 University Ave W, Waterloo, ON N2L 3G1, Canada}

\author[0000-0001-7145-8674]{F.~Prada}
\affiliation{Instituto de Astrof\'{i}sica de Andaluc\'{i}a (CSIC), Glorieta de la Astronom\'{i}a, s/n, E-18008 Granada, Spain}

\author[0000-0001-6979-0125]{I.~P\'erez-R\`afols}
\affiliation{Departament de F\'isica, EEBE, Universitat Polit\`ecnica de Catalunya, c/Eduard Maristany 10, 08930 Barcelona, Spain}

\author{G.~Rossi}
\affiliation{Department of Physics and Astronomy, Sejong University, 209 Neungdong-ro, Gwangjin-gu, Seoul 05006, Republic of Korea}

\author[0000-0002-9646-8198]{E.~Sanchez}
\affiliation{CIEMAT, Avenida Complutense 40, E-28040 Madrid, Spain}

\author{D.~Schlegel}
\affiliation{Lawrence Berkeley National Laboratory, 1 Cyclotron Road, Berkeley, CA 94720, USA}

\author[0000-0002-3461-0320]{J.~Silber}
\affiliation{Lawrence Berkeley National Laboratory, 1 Cyclotron Road, Berkeley, CA 94720, USA}

\author{D.~Sprayberry}
\affiliation{NSF NOIRLab, 950 N. Cherry Ave., Tucson, AZ 85719, USA}

\author[0000-0003-1704-0781]{G.~Tarl\'{e}}
\affiliation{University of Michigan, 500 S. State Street, Ann Arbor, MI 48109, USA}

\author{B.~A.~Weaver}
\affiliation{NSF NOIRLab, 950 N. Cherry Ave., Tucson, AZ 85719, USA}

\author[0000-0001-5381-4372]{R.~Zhou}
\affiliation{Lawrence Berkeley National Laboratory, 1 Cyclotron Road, Berkeley, CA 94720, USA}

\author[0000-0002-6684-3997]{H.~Zou}
\affiliation{National Astronomical Observatories, Chinese Academy of Sciences, A20 Datun Road, Chaoyang District, Beijing, 100101, P.~R.~China}

\collaboration{50}{(DESI Collaboration)}

\received{2026 March 11}
\revised{2026 May 18}
\accepted{2026 May 19}
\published{2026 July 6}

\begin{abstract}
Stellar streams are the result of a host galaxy’s gravitational potential tidally disrupting satellite dwarf galaxies and globular clusters (GCs), causing them to grow leading and trailing tidal tails. The C-19 stellar stream is an extremely metal-poor stellar population, showing chemical abundance patterns characteristic of a GC. However, its large velocity dispersion is difficult to reconcile with a conventional, purely baryonic, disrupting-GC progenitor. Current techniques for stream characterization are primarily applied to \textit{Gaia} DR3, relying heavily on proper motion measurements. Using the Dark Energy Spectroscopic Instrument (DESI), which provides radial velocities and metallicities for over 10 million stars reaching significantly fainter magnitudes than comparable surveys, we employ a mixture model approach to jointly characterize stream populations in proper motions, radial velocities, and metallicities against a Milky Way halo background. By applying this framework to the C-19 stellar stream, we identify a total of 47 spectroscopically confirmed member stars, of which 41 are newly identified and only 6 were previously reported in the literature. In this work, we measure a velocity dispersion of $7.8^{+1.5}_{-1.3}$ km s$^{-1}$ and a mean metallicity of [Fe/H] = $-3.36^{+0.12}_{-0.10}$. We further identify a novel `spur’ feature within the stream. We conclude that our measurements are in line with previous works identifying C-19 as a `hot', metal-poor stream. In forthcoming work, we will apply this approach to many more streams in the DESI footprint, enabling population-level comparisons with predictions from simulations.
\end{abstract}

\section{Introduction} \label{sec:intro}

The modern hierarchical model of galaxy evolution suggests that Milky Way (MW)-like galaxies formed through mergers of dwarf galaxies (DGs) and globular clusters (GCs) \citep{white_core_1978, johnston_interpreting_2001, bullock_tracing_2005}. An intermediate step in the accretion of DGs and GCs is the formation of tidal tails---either through shocks at the pericenter of their orbit or through gradual removal of stars across the tidal radius---before eventually phase-mixing into the host galaxy's population \citep{lynden-bell_ghostly_1995, oh_tidal_1995}. Stellar streams are the present-day occurrences of this intermediate step; through their study, we can improve our understanding of the earliest formed stellar populations \citep{belokurov_discovery_2006, belokurov_field_2006, casey_signature_2021,martin_stellar_2022,tavangar_fire_2022, yuan_pristine_2022,koposov_s_2023,yuan_pristine_2025, bayer_pristine-unions_2025}, the Milky Way's potential \citep{ibata_great_2001,helmi_velocity_2004, johnston_two_2005, law_sagittarius_2010,KoposovSE2010a, bovy_shape_2016}, and the dark matter substructure within our Galaxy \citep{ibata_uncovering_2002,johnston_how_2002,siegal-gaskins_signatures_2008,erkal_forensics_2015,ngan_dispersal_2016,sanders_dynamics_2016,erkal_sharper_2017,  bonaca_spur_2019, nibauer_textttstreamsculptor_2024, lu_detectability_2025}.

The number of known stellar streams identified within the MW has reached $\sim$ 100 thanks to surveys of our stellar halo (e.g., SDSS, \citey{york_sloan_2000}; \textit{Gaia}, \citey{gaia_collaboration_gaia_2016}; DES, \citey{dark_energy_survey_collaboration_dark_2016}) and stream detection algorithms (e.g., matched filter, \citealt{rockosi_matched-filter_2002, shipp_stellar_2018}; co-moving group detection, \citey{williams_dawning_2011}; pole counts, \citey{johnston_fossil_1996}; \texttt{STREAMFINDER}, \citealt{malhan_streamfinder_2018, ibata_charting_2021, ibata_charting_2023}). Furthermore, the number of identified stellar streams is expected to increase dramatically over the next decade with the advent of the Vera C. Rubin Observatory's Large Survey of Space and Time (LSST) \citep{lsst_science_collaboration_lsst_2009}. The LSST will observe stars down to magnitudes of $r \sim 27$ \citep{lsst_science_collaboration_science-driven_2017}, representing a significant improvement over current surveys \citep{drlica-wagner_probing_2019}.

The C-19 stellar stream merits study in its own right. Discovered by the \texttt{STREAMFINDER} (hereafter, SF) algorithm \citep{ibata_charting_2021}, applied to the \textit{Gaia} Data Release 3 \citep{gaia_collaboration_gaia_2021}, C-19 is the most metal-poor stellar population discovered to date \citep{martin_stellar_2022}. Analysis of its chemistry shows a mean metallicity of [Fe/H] $< -3.0$ with a spread less than 0.2 dex. Prior to confirmations of the Phoenix \citep{wan_tidal_2020} and C-19 \citep{martin_stellar_2022} stellar streams, a `metallicity floor' of [Fe/H] $\gtrsim -2.5$ was thought to be universal for stellar clusters \citep{beasley_old_2019}. 

C-19's variations in light element abundances \citep{martin_stellar_2022} are characteristic of a GC \citep{gratton_abundance_2004, gratton_what_2019}.  However, the stream velocity dispersion of identified member-stars is hotter than most known GC streams ($\sim2-5 \text{ km s}^{-1}$, see \citey{li_s_2022}). Initial measurements of C-19's velocity dispersion \citep{martin_stellar_2022}, based on eight stars spanning $15{^\circ}$ along the stream, found a dispersion in line-of-sight velocities of $\sim 7$ km s$^{-1}$. More recent measurements, using 23 stars with radial velocities spanning $\sim 100{^\circ}$, derived a velocity dispersion of $11.1_{-1.6}^{+1.9}$ km s$^{-1}$ \citep{yuan_pristine_2025}. These values are in tension with a GC origin, whose velocity dispersions typically fall between 2 - 5 km s$^{-1}$ \citep{li_s_2022}.

This tension may provide insight into the dark matter substructure of the MW halo. In a smooth galactic potential, simulations predict dynamically cold, spatially thin streams from GC progenitors \citep{ibata_uncovering_2002,yoon_clumpy_2011,Carlberg25_C19}. However, cold dark matter (CDM) models predict a large number of subhalos, many of which do not form a stellar component at all \citep{bullock_reionization_2000,benson_effects_2002,diemand_formation_2007,springel_aquarius_2008,jethwa_upper_2018, nadler_impact_2025}. \citet{Carlberg25_C19} investigates the `heating' of C-19, where gravitational interactions with dark subhalos perturb the stellar stream, increasing the velocity dispersion. Alternatively, C-19 may have first experienced heating within a parent DG before later merging with the MW; the current kinematics of C-19 could probe the central DM density profile of its parent dwarf galaxy \citep{malhan_butterfly_2019,bonaca_orbital_2021,malhan_new_2022}. It is also feasible that a dark matter dominated DG progenitor, such as an ultra-faint dwarf, could allow for the kinematic properties of C-19 \citep{errani_structure_2022,errani_pristine_2022}. 

In this paper, we provide our own analysis of the C-19 stellar stream using data from the first three years of the Dark Energy Spectroscopic Instrument (DESI) \citep{prieto_preliminary_2020,cooper_overview_2023, koposov_desi_2024}. DESI's Milky Way Survey (MWS) pipeline provides radial velocities, stellar parameters, and metallicities for orders of magnitude more stars than existing spectroscopic surveys with similar resolutions. The instrument is well equipped to do large-scale stellar spectroscopy, and has observed approximately 15 million stars within our MW to a magnitude of $r = 19$ after 3 years of observations. As a result, the DESI footprint covers $\sim 100$ known stellar streams \citep{mateu_fourteen_2018, ibata_charting_2023, bonaca_stellar_2024}.

In this work, we primarily rely on mixture modelling to identify stars that are members of the stellar stream from those in the `background' (which we define as any star not part of the stream) \citep{rasmussen_infinite_1999,koposov_piercing_2019,kuzma_searching_2020,kuzma_detecting_2021}. We work with the assumption that the data in the field about a stellar stream can be well described by two distinct stellar populations: that of the stellar stream and the background.

Mixture modelling offers several advantages over other clustering methods (e.g., K-means, \citealt{lloyd_least_1982}; DBSCAN, \citealt{ester_density-based_1996}), including scalability for many populations, the incorporation of measurement errors, less sensitivity to outliers, memory efficiency, and the ability to assign membership probabilities to stars for more nuanced analysis \citep{xu_comprehensive_2015}. 

In this paper, we use 3 years of DESI data to study the C-19 stellar stream, identifying 41 new spectroscopic members through modelling the kinematics and chemistry of the stream. We find that the C-19 stellar stream is kinematically hot and extremely metal poor, in alignment with past studies of C-19 \citep{martin_stellar_2022, yuan_pristine_2022, yuan_pristine_2025}.
This paper is structured as follows. \S\ref{sec:data} describes the data we use in our analysis. In \S\ref{sec:methods} we outline the mixture model used to identify high-probability C-19 members. \S\ref{sec:results} presents the resulting kinematic, chemical, and morphological properties of the stream, alongside validation of our method and assumptions. In \S\ref{sec:orbit} we model C-19’s orbit. \S\ref{sec:discussion} examines the influence of the MWS selection function and discusses what we can infer regarding the progenitor's properties. We conclude in \S\ref{sec:conclusion}.

\section{Data} \label{sec:data}

\subsection{DESI Milky Way Survey}

DESI is a new-generation multi-object survey spectrographic facility operating on the Mayall 4-meter telescope at Kitt Peak National Observatory \citep{,abareshi_overview_2022}. It uses 5,000 fibers for collecting data over a $3.2{^\circ}$ diameter field of view and a wavelength range of 360 to 980 nm (R $\sim$ 2000 to 5500) \citep{collaboration_desi_2016,guy_spectroscopic_2023,schlafly_survey_2023, miller_optical_2024, poppett_overview_2024}. We use data from the first three years of observations, which are scheduled as part of the 2nd public data release for DESI in 2027 (hereafter DR2). The main mission of DESI is to constrain the properties of dark energy leveraging the most precise measurements of the expansion history of the universe to date \citep{levi_desi_2013}. The first data release (DR1, \citealt{desi_collaboration_data_2025}) includes spectra for over 18 million unique targets, and early results from DESI have placed new constraints on dark energy \citep{adame_desi_2025, desi_collaboration_desi_2025}.

Our work relies on DESI observations of our own Galaxy, largely comprised of data collected as a part of the MWS program. The target selection, observation strategy, and data pipeline of the MWS is described in detail across \citet{prieto_preliminary_2020}, \citet{cooper_overview_2023}, \citet{koposov_desi_2024, KoposovSE2025}, and Koposov et al. (2026, in prep.). The MWS observes stars during DESI's Bright Time Survey; dark conditions with good transparency is reserved for the galaxy and quasar redshift survey. The Bright Time Survey's fibre assignment, prioritization, and observing time are shared between the Bright Galaxies Survey and the MWS. During this time, the MWS observes a magnitude limited selection between $16 < r < 19$. 

The MWS is complemented by the Milky Way Backup Program (MWBP), which was operated in twilight or poor weather \citep{dey_backup_2025}. The MWBP includes fainter sources ($G < 19$ mag) and unbiased sample of bright stars ($11.2 < G < 16$ mag), significantly extending DESI's magnitude range for targets within our Galaxy. The DESI DR1 stellar catalogue \citep{KoposovSE2025} describes the entire stellar sample across all survey programs; going forward in this work, we refer to the stellar catalogue in its entirety as the MWS.



\begin{figure*}[!ht]
    \centering
    \includegraphics[width=0.8\linewidth]{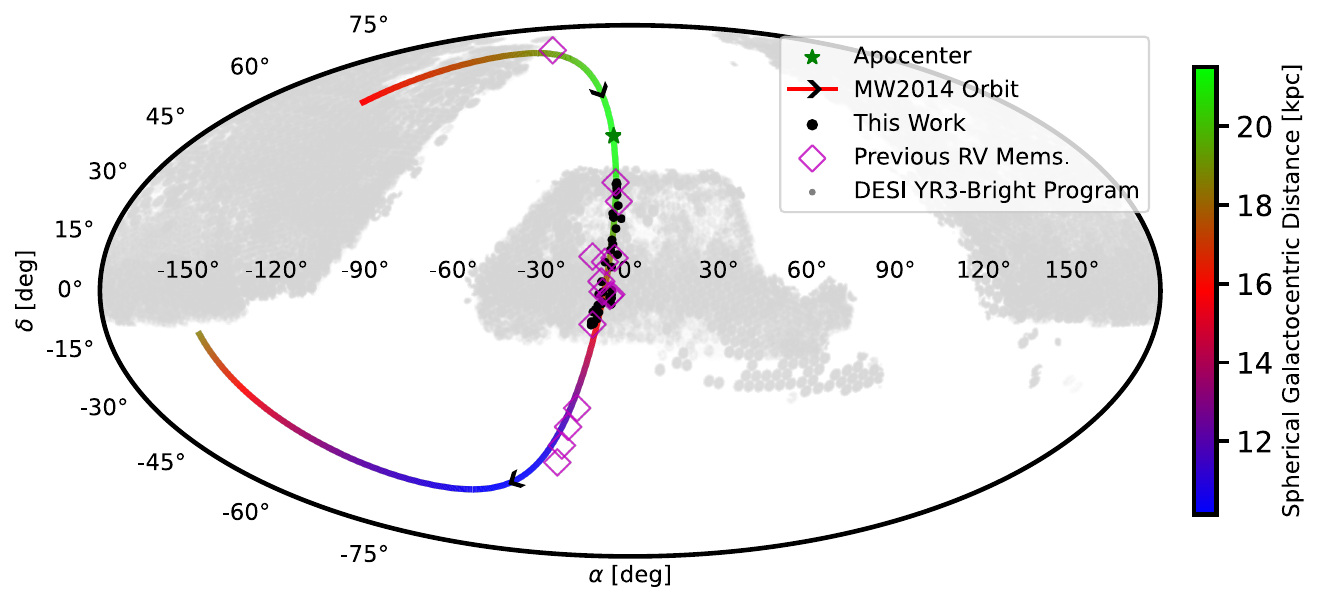}
    \caption{The black points represent the 47 (41 main sequence + red giant, 6 blue horizontal branch) C-19 member candidates identified in this work. We mark the previously known C-19 member stars that have radial velocity measurements with magenta diamonds.  The DESI MWS's footprint from 3 years of observations is shown in grey. We also visualize our fitted C-19 progenitor orbit in \texttt{MWPotential2014} (\S \ref{sec:orbit}; \citey{bovy_galpy_2015}), coloured by the galactocentric distance in kpc. The orbit direction is indicated with arrows along its path.}
    \label{fig:footprint}
\end{figure*}

Using \texttt{RVSpecfit} \citep{koposov_rvspecfit_2019}, the MWS fits the radial velocity and stellar parameters for all of its targets. The pipeline relies on synthetic spectra from the PHOENIX grid \citep{husser_new_2013}. We convert the radial velocities to the Galactic Standard of Rest (GSR) reference frame ($v_\text{GSR}$) using \texttt{astropy}'s ICRS frame, with $v_\odot$ =(-12.9, 245.6, 7.78) km s$^{-1}$ \citep{drimmel_solar_2018}. 

Later, for verification of our work, we use spectrophotometric distances inferred using the \texttt{rvsdistnn} machine learning method described in more detail in \citet{aganze_cocytos_2025} and Koposov et al. (2026, in prep.)\footnote{We note that DESI has provided a distance catalog for DR1 \citep[\texttt{SpecDis;}][]{li_specdis_2025}, which is derived directly from DESI spectra, rather than from stellar parameters as used in this work.}. \texttt{rvsdistnn} maps stellar parameters ($\log g, T_\text{eff}, [\text{Fe/H}], [\alpha/\text{Fe}]$) derived by \texttt{RVSpecFit} \citep{koposov_rvspecfit_2019} for the 15 million stars observed into absolute magnitudes, from which heliocentric distances are computed via the distance modulus. 
Since uncertainties are naturally quantified in magnitude units (i.e., in distance modulus), all distance plots in this paper are shown on a logarithmic scale. Furthermore, given the relatively large uncertainties ($\sim$10--20\%), the derived heliocentric distances are used only for visualization and validation purposes and are not incorporated into our mixture model.

We also use the blue horizontal branch (BHB) catalogue amassed by \citet{bystrom_exploring_2024}. Within the DESI MWS, BHBs are assigned high targeting priority; along with other classes such as white dwarfs and nearby stars, they are preferentially allocated fibres to ensure high completeness.

\subsection{Ancillary Data}
We cross-match the MWS DR2 with data from the Dark Energy Camera Legacy Survey (DECaLS) Data Release 9 for the photometric properties of stars \citep{dey_overview_2019}, and with the \textit{Gaia} Data Release 3 (\textit{Gaia} DR3; \citey{gaia_collaboration_gaia_2023}) for astrometric measurements.
\subsubsection{Dark Energy Camera Legacy Survey}

DECaLS is one of the three individual surveys that, in complement, constitute the DESI Legacy Imaging Surveys (hereafter the Legacy Surveys, \citey{dey_overview_2019}) used for optical targeting of DESI. DECaLS uses the Dark Energy Camera (DECam; \citey{flaugher_dark_2015}) at the 4-m Blanco telescope in Chile. We use the DECaLS photometric data in the de-reddened $grz$ bands in this work to construct colour-magnitude and colour-colour selections. The reddening correction was performed using the extinction coefficients taken from the DES Data Release 1 \citep{abbott_dark_2018} and the \citet{schlegel_maps_1998} $E(B-V)$ values.



\subsubsection{\textit{Gaia} Data Release 3}

\textit{Gaia} Data Release 3 (\textit{Gaia} DR3;\citey{gaia_collaboration_gaia_2023}) is the third full data release from the European Space Agency's \textit{Gaia} mission \citep{gaia_collaboration_gaia_2016}. \textit{Gaia} DR3 provides parallax and proper motions for 1.5 billion sources to a magnitude of G $\lesssim$ 21 (\citey{gaia_collaboration_gaia_2021}). We use the proper motions in right ascension, $\mu_\alpha \cos \delta$ (hereafter simply $\mu_\alpha$), and declination, $\mu_\delta$, in modelling the kinematics of the C-19 stellar stream.

\subsection{Data Preparation}\label{sec:prep}

We apply several quality cuts to our combined catalogue. First, we require one observation per object and restrict the sample to sources classified as stars by \texttt{REDROCK} (i.e. \texttt{RR\_SPECTYPE} $=$ \texttt{STAR}) \citep{anand_archetype-based_2024}. We remove measurements with large radial-velocity and metallicity uncertainties ($v_\text{err}>10~\mathrm{km \ s^{-1}}, \text{[Fe/H]}{_\text{err}} > 0.5 \text{ dex}$) and exclude rows with non-zero \texttt{RVSpecFit} warnings (i.e. \texttt{RVS\_WARN $\neq$ 0}). Finally, we discard all rows with missing entries in any of the following fields: $E(B-V)$ colour excess, $grz$ photometry, parallax $\varpi$, and proper motions $\mu_\alpha, \mu_\delta$. We corrected for underestimated uncertainties in radial velocities and proper motions by adding in quadrature systematic uncertainty floors of 1.0 km s$^{-1}$ \citep{KoposovSE2025} and $\sqrt{550}\times10^{-3} \text{ mas yr}^{-1}$ \citep{gaia_collaboration_gaia_2021}, respectively. 

\subsection{Initial Sample Selection} \label{sec:trunc}

To aid the performance of our mixture model, we truncate our dataset within some tolerance about the known literature measurements of the C-19 stream properties, increasing the fraction of stars in our field belonging to the stream. To inform this selection, we consider C-19 member stars that have previously been identified by the SF algorithm applied to \textit{Gaia} DR3. SF is a generic algorithm that aims to work with any mix of data sets for any combination of positions and kinematics to identify stellar streams \citep{malhan_streamfinder_2018}. When applied to \textit{Gaia} DR3, SF identifies 46 stars as members; 12 of these stars had measured radial velocities \citep{ibata_charting_2023}. Six of the SF members have been observed in the MWS DR2; we can use the DESI measurements of these stars to inform cuts.

We use a great-circle coordinate transform to rotate the data into celestial stream coordinates ($\alpha$, $\delta$, to $\phi_1$, $\phi_2$) defined for C-19 in \citet{ibata_charting_2023} as  $\alpha_0 =  354.356^\circ, \alpha_\mathrm{pole} =  81.45^\circ, \delta_\mathrm{pole}=-6.346^\circ$. Here, $\alpha_0$ sets the $\phi_1=0$ zero-point, and $(\alpha_\mathrm{pole}, \delta_\mathrm{pole})$ are the right ascension and declination of the stream-frame pole. This rotation places the stream's track roughly parallel to $\phi_2 = 0$. We keep stars with $ \phi_2  \in [-5, 5]$, while our $\phi_1$ bounds are placed by the footprint of the MWS DR2 data ($\phi_1 = -8^\circ$: Galactic plane, $\phi_1 = +38^\circ$: southern hemisphere) as shown in Figure \ref{fig:footprint}.  We find that increasing the $\phi_2$ bounds does not change our findings in this work. 

\citet{martin_stellar_2022} finds C-19 to be at a distance of 18 kpc with no significant distance gradient (also refer to \S \ref{sec:orbit}). We remove foreground stars nearer than 18 kpc from consideration based on their \textit{Gaia} parallax using
\begin{equation}\label{eq:plx}
    \varpi - 2\sigma_{\varpi} > \frac{1}{18 \text{ kpc}}.
\end{equation}

We also apply a colour-magnitude cut defined relative to a model sequence and red giant branch (MS+RGB) isochrone obtained from \citet{dotter_mesa_2016}. We adopt the lowest metallicity isochrone available, [Fe/H] $=-2.49$ dex, and select the associated age to be $13.5$ Gyrs. Stars are kept if they lie within $\delta(g-r) = |(g-r)-(g-r)_{\text{iso}}| \leq 0.2$ of the isochrone at a given magnitude. Due to large uncertainties in the metallicities of horizontal branch stars, we consider them separately in \S \ref{sec:box-cuts}. 

Despite the metallicity of C-19 being nearly a dex below this isochrone, we expect this to have little to no impact on our findings as the leftward shift of the red giant branch on the colour-magnitude diagram becomes increasingly small at such low metallicities. Moreover, our padding about the isochrone serves as an additional safeguard for this potential discrepancy.

\begin{figure}
    \centering
    \includegraphics[width=1\linewidth]{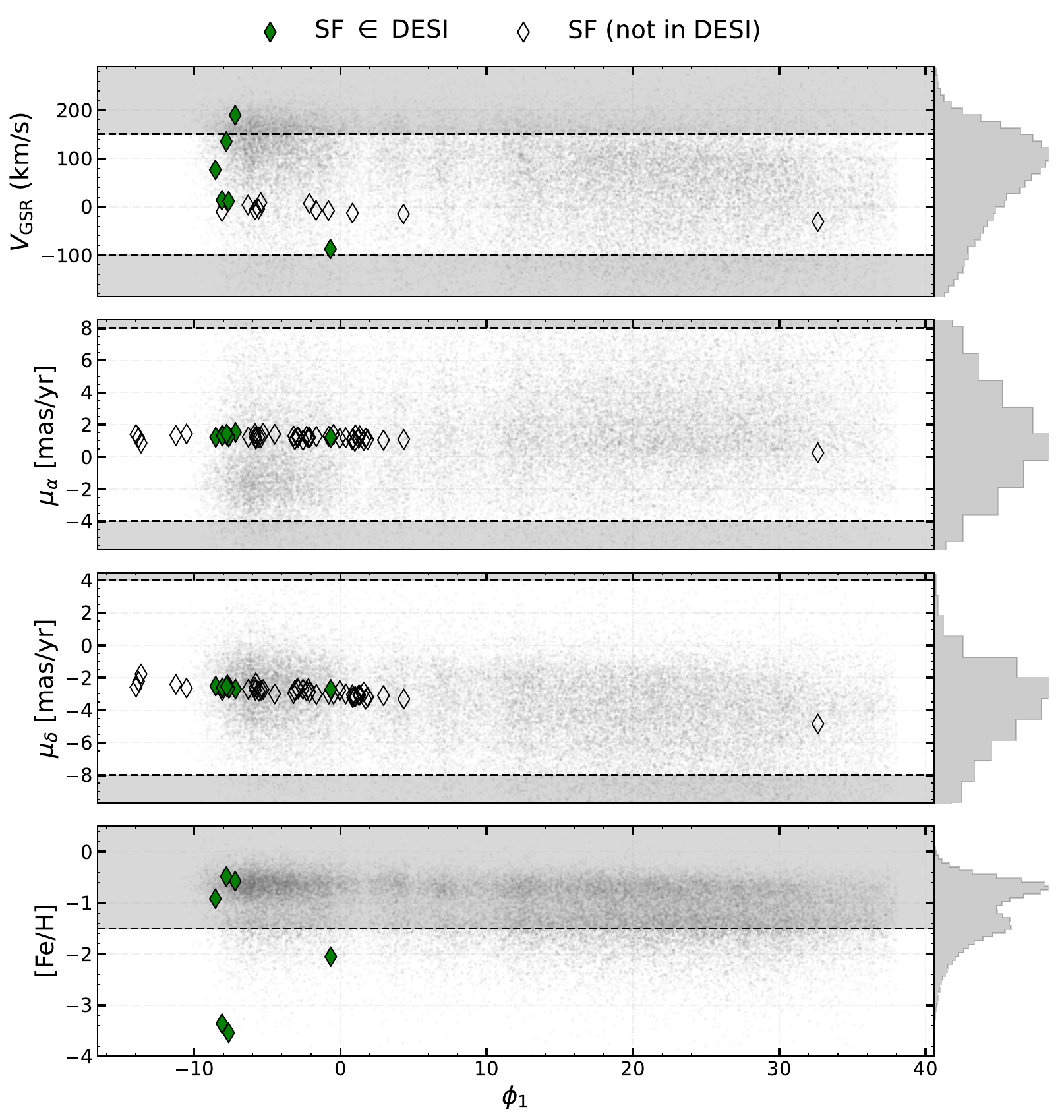}
    \caption{Chemical and kinematic measurements versus on-sky position ($\phi_1$)  for the parameters in the mixture model. We make truncations in these parameters in order to increase the fraction of stream members with respect to the surrounding field; truncated regions are shown in grey. The truncations are informed by the \texttt{STREAMFINDER} (SF; \citealt{ibata_charting_2021, ibata_charting_2023}) \textit{Gaia} DR3 members. The SF members' proper motion values are taken from \textit{Gaia} DR3, while line-of-sight velocity data either comes from \textit{Gaia} DR3 or from spectroscopic follow up with the VLT/UVES spectrograph in \citet{ibata_charting_2023}. Those stars also observed in DESI DR2 are shown as filled green diamonds.}
    \label{fig:trunc}
\end{figure}

The next cuts we describe are about our model parameters: [Fe/H], $v_\text{GSR}, \ \mu_\alpha, \ \mu_\delta$ ; we account for these cuts later by implementing truncated-Gaussians for the background population in our mixture model. The cuts made are as follows:
\begin{itemize}[label={-}]
    \item $v_{\rm GSR} \in [-100,\,150]~{\rm km\,s^{-1}}$
    \item $\mu_{\alpha} \in [-4,\,8]~{\rm mas\,yr^{-1}}$
    \item $\mu_{\delta} \in [-8,\,4]~{\rm mas\,yr^{-1}}$
    \item $[{\rm Fe}/{\rm H}] < -1.5~{\rm dex}$
\end{itemize}
The cuts are visualized by the shaded regions in Figure \ref{fig:trunc}, and are made to both include any distinct trends made by the SF members and enough of the background population to accurately characterize it. We are left with 2,071 stars in our dataset. It is notable that only two of the six SF members observed by DESI survive the joint selection. Although they all agree with the kinematic track in proper motion space, radial velocities and metallicities from DESI aid discrimination between members and contaminants.

\section{Mixture model}\label{sec:methods}
\subsection{Stream and Background models}
We construct a two component mixture model for the observables $x =  (v_{\rm GSR},\mu_\alpha,\mu_\delta, \rm [Fe/H])$ given position along the stream $\phi_1$. Each component is normally distributed about some mean $(E_x)$ with an intrinsic dispersion $(\sigma_x)$. We choose not to fit a stream track for $\phi_2$ as not to assume an on-sky shape or a constant stream width; otherwise, we may miss possible spur, kink, or under-density features within the stream. We treat the background means as constant across our field. However, to reflect the fact that the stream members approximate the orbit of their progenitor we allow the stream's $E_{v_\text{GSR}}, E_{\mu_\alpha}, \text{and } E_{\mu_\delta}$ to vary with position along the stream $\phi_1$. To achieve this, we model their means as cubic splines with subintervals set by $N$ equally spaced knots along the stream's extent. For example, let $\phi_{1,i}$, for $i=1,\dots,N$, denote the knots' positions along the stream. Then, the stream's expectation value for $v_\text{GSR}$ is given by:

\begin{equation}
    E_{v_{\rm GSR}}(\phi_1) = \begin{cases}
        P_1(\phi_1), & \phi_{1, 1} \le \phi_1<\phi_{1, 2} \\
        P_2(\phi_1), & \phi_{1, 2} \le \phi_1<\phi_{1, 3} \\
        ... \\
        P_{N-1}(\phi_1), & \phi_{1, N-1} \le \phi_1 < \phi_{1, N},
    \end{cases}
\end{equation}
where each $P_i(\phi_1)$ is a cubic polynomial defined on the interval between knot $
\phi_{1, i}$ and $\phi_{1, i+1}$, and the value of each knot is $E_{v_\text{GSR}, i} = E_{v_\text{GSR}}(\phi_{1, i})$. In contrast to the kinematics of the stream, we treat its mean metallicity, $E_{\text{[Fe/H]}}$, as a constant in $\phi_1$.

To model C-19, we adopt $N=5$ equally spaced spline nodes beginning at $\phi_1 = -9^\circ$ and ending at $\phi_1=38^\circ$. Finally, the full model is parameterized as $\theta_{\rm tot} = \{\theta_{\rm st};\,\theta_{\rm bg}\}$, with
\begin{equation}
\begin{split}
\theta_{\rm st}= \big\{[E_{x, 1} ... E_{x, 5}]_{x \neq \text{[Fe/H]}}, E_{\rm [Fe/H]},  \sigma_{x}\big\} ,
     \\
     \theta_{\rm bg} = \big\{E_{x}, \sigma_{x}\big\}.
\end{split}
\end{equation}

\subsection{Likelihood and prior}

We operate in a Bayesian framework to find the best-fit model to discern the C-19 stellar stream from the background population. This technique allows us to naturally incorporate measurement uncertainties in our likelihood and the current literature values into our prior distribution.  Our approach is similar to the approach constructed in \citet{awad_s5_2025}; our aim is to find the posterior distribution according to Bayes' Theorem:

\begin{equation}
p(\theta|x) \propto {{p}(x|\theta) \cdot p(\theta)},
\end{equation}
where for the observed data $x$, the posterior is proportional to the product of the likelihood of the data and the prior probability of the model. 

To account for measurement uncertainties, we define the total variance for each observable as $\zeta_{k}^2 = \sigma_{x}^2 + \delta_{k}^2$, where $\sigma_{x}$ is the fitted intrinsic dispersion, and $\delta_{k}$ is the reported measurement error for the $k$-th star. 

The total likelihood function for our mixture model is
\begin{equation}
    {p}(x|\theta_\text{tot}, \phi_1) =  f_s\cdot{p}(x 
|\theta_\text{st}, \phi_{1}) + (1-f_s) \cdot {p}(x|\theta_\text{bg} ),
\end{equation}
where $f_s$ is the fraction of stars in the sample that belong to the stream.

The stream likelihood is the product over the $N$ stars and the four independent stream expected values:
\begin{equation}
    {p}(x|\theta_\text{st}, \phi_1)  =  \prod _{k=1}^{N} \prod_{i=1}^{4} \frac{1}{\zeta_{k,i}\sqrt{2 \pi}} \text{ exp} \left\{\frac{-\left[x_{k,i}-E_{x_{\text{i}}}  (\phi_{1, k}) \right]^2}{2 \zeta_{k,i}^2} \right\}.
\end{equation}
The background likelihood is expressed similarly, except we must re-normalize the likelihood due to the data truncation done in \S \ref{sec:trunc} by dividing by the probability mass:

\begin{equation}
       {p}(x|\theta_\text{bg}) = \prod _{k=1}^{N} \prod_{i=1}^{4} \frac{\frac{1}{\zeta_{k,i}\sqrt{2 \pi}} \text{ exp} \left\{\frac{-\left[x_{k,i}-E_{x_{\text{bg,i}}} \right]^2}{2 \zeta_{k,i}^2} \right\}}{p(a< x \leq b)},
\end{equation}
where $p(a< x \leq b) = F_x(b) - F_x(a)$ is the difference between the normal cumulative distribution function (CDF) evaluated at the upper and lower truncation bounds $(b, a)$. To first order, we do not need to re-normalize the stream likelihood because the truncations occur multiple standard-deviations away from the stream's mean. In practice, we implement the likelihood and CDF calculations in log space, using the \texttt{stats.norm.logpdf} and \texttt{stats.norm.logcdf} from the \texttt{scipy} python package, to improve numerical stability and efficiency in the computation.

To infer a membership probability, $P_{k \ \in \text{ stream}}$, between 0\% and 100\% of each star belonging to the stream population, we calculate the membership probability for each star as
\begin{equation} \label{eq:P}
    P_{\ \in\text{ stream}}  = \frac{f_s\cdot{p}(x |\theta_\text{stream}, \phi_{1})}{{p}(x|\theta_\text{tot}, \phi_{1})}\times 100\%,
\end{equation}
where $f$ is the fraction of stars belonging to the stream, and $p(x|\theta, \phi_1)$ is the likelihood of the observed data $x$ given the inferred model parameters $\theta$ and position along the stream $\phi_1$.

We incorporate a uniform prior for the component means and  stream fraction, and $\log$-uniform prior for the dispersions. Our prior ranges are set by visual inspection of the total distribution; the ranges are listed in Appendix \ref{app:prior}.

\subsection{MCMC} \label{sec:mcmc}

The sampling of the posterior likelihood is undertaken with \texttt{emcee}, a python implementation of the Markov Chain Monte Carlo (MCMC) method by \citet{foreman-mackey_emcee_2013}. We use 70 walkers, running 50,000 steps and discard the first 6,000 as burn-in. We then thin to reduce autocorrelation in the final chains, retaining every other sample. In total, we fit 27 posterior dimensions between the stream and background components, listed in Table \ref{tab:fitted}. We show subsets of the corner plot in Appendix \ref{app:ppc}. 

We find through testing that our model is unable to constrain the intrinsic dispersions in the stream proper motions, and the posterior is more or less uniformly distributed within our prior range. Therefore, we fix $\sigma_{\mu_\alpha},\sigma_{\mu_\alpha} = 0.09$ mas yr$^{-1}$, equivalent to a velocity dispersion $\sim 8$ km s$^{-1}$ at the literature value of the C-19 heliocentric distance (i.e. 18 kpc). This assumes that the kinematic dispersion of the stream has no preferential direction, and as such, the dispersion in the plane of the sky would be similar to the value we converge on for $\sigma_{v_\text{GSR}}$. Given the average proper motion uncertainties of the stream members are around 0.2 mas\,yr$^{-1}$, it is not surprising that the intrinsic dispersion in stream proper motion can not be constrained when it is at $<0.1$\,mas yr$^{-1}$ level.

Additionally, we run a 35-parameter version of our model where we allow both the stream fraction $f_s$ and the velocity dispersion $\sigma_{v_\text{GSR}}$ to vary along $\phi_1$ using 5 spline nodes. When we calculate membership probability $P_{\in \text{stream}}$ from this model we retrieve the same member stars above the $50\%$ threshold. We used the varying tracks of $f_s$  and $\sigma_{v_\text{GSR}}$ in \S \ref{sec:verify} to validate our model's assumptions, and in Figure \ref{fig:vdisp} to study how the kinematics vary along the length of the stream.

\begin{table}
\centering
\caption{Results from the MCMC run in section \ref{sec:mcmc}; 27 parameters were fitted in total using 2,071 stars in the MWS DR2. 
The $E_{v_{\mathrm{GSR}}, i}, E_{\mu_{\alpha}, i}, \text{and } E_{\mu_{\delta}, i}$ parameters represent the 5 spline nodes equally spaced along the stream length between $\phi_1 = -9^{\circ}$ and $38^{\circ}$; their values are given in Table \ref{tab:spline}.}
\begin{tabular*}{0.9\linewidth}{@{\extracolsep{\fill}} l r}\label{tab:fitted}

Parameter & Value \\
\hline
\multicolumn{2}{l}{\small{\textit{Stream component}}} \\[0.5pt]

$f_s$ (\%) & $2.0_{-0.3}^{+0.3}$ \\

$\sigma_{v_\mathrm{GSR}}$ (km\,s$^{-1}$) 
  & $7.8_{-1.3}^{+1.5}$ \\[3pt]

$E_{\mathrm{[Fe/H]}}$ (dex) & $-3.36_{-0.10}^{+0.12}$ \\
$\sigma_{\mathrm{[Fe/H]}}$ (dex) 
  & $0.23_{-0.04}^{+0.05}$ \\[3pt]

$E_{v_{\mathrm{GSR}},i}$ (km\,s$^{-1}$) & see Table \ref{tab:spline} \\
$E_{\mu_{\alpha,i}}$ (mas\,yr$^{-1}$) & see Table \ref{tab:spline} \\
$E_{\mu_{\delta,i}}$ (mas\,yr$^{-1}$) & see Table \ref{tab:spline} \\

\hline

\multicolumn{2}{l}{\small{\textit{Background component}}} \\[0.5pt]

$E_{v_{\text{GSR}}}$ (km\,s$^{-1}$) & $15.4_{-7.2}^{+9.5}$ \\
$\sigma_{v_{\text{GSR}}}$ (km\,s$^{-1}$) 
  & $128.4_{-9.9}^{+12.6}$ \\[3pt]

$E_{\mathrm{[Fe/H]}}$ (dex) & $-0.79_{-0.04}^{+0.04}$ \\
$\sigma_{\mathrm{[Fe/H]}}$ (dex) 
  & $0.83_{-0.07}^{+0.09}$ \\[3pt]

$E_{\mu_{\alpha}}$ (mas\,yr$^{-1}$) & $2.17_{-0.06}^{+0.06}$ \\
$\sigma_{\mu_{\alpha}}$ (mas\,yr$^{-1}$) 
  & $2.61_{-0.06}^{+0.06}$ \\[3pt]

$E_{\mu_{\delta}}$ (mas\,yr$^{-1}$) & $-3.66_{-0.06}^{+0.06}$ \\
$\sigma_{\mu_{\delta}}$ (mas\,yr$^{-1}$) 
  & $2.43_{-0.05}^{+0.05}$ \\
\hline
\end{tabular*}

\vspace{4pt}
\end{table}

\section{Results}\label{sec:results}
\subsection{High-Probable Members}

We assign a membership probability to each of the 2,071 stars by drawing 1,000 samples from our posterior distribution and computing $P_{\in \rm  stream}$ using Equation \ref{eq:P}. The final membership probability for a given star is taken to be the median of its resulting $P_{\in \rm  stream}$ values. Uncertainties are derived from the $16$th and $84$th percentiles of the posterior distribution, with the lower and upper error bars given by the difference between the median and the $16$th and $84$th percentiles, respectively, representing a $68\%$ credible interval. Table \ref{tab:members} lists all stars with membership probabilities above $0.1\%$.

\begin{figure}
    \centering
    \includegraphics[width=1\linewidth]{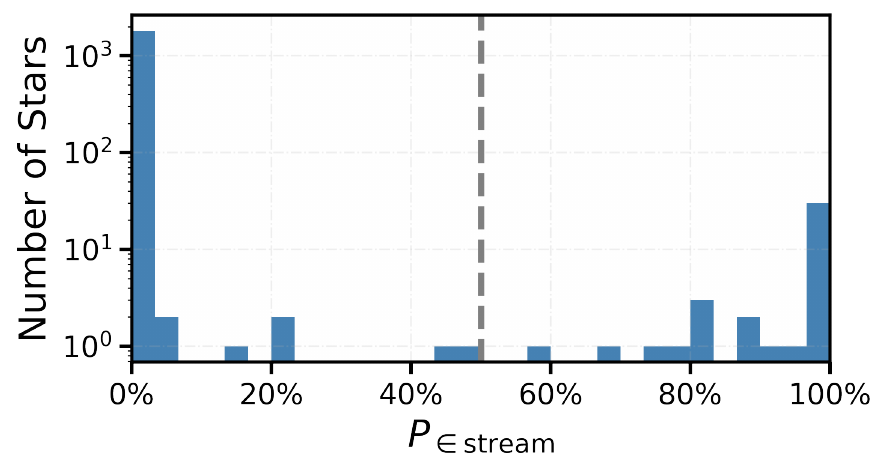}
    \caption{Histogram showing the distribution of stream membership probability ($P_{\in \text{ stream}}$) for all 2071 stars modelled. The 41 stars with membership probability greater than 50\% (grey line) are considered member stars in this work.}
    \label{fig:memhist}
\end{figure}

\begin{figure*}
    \centering
    \includegraphics[width=1\linewidth]{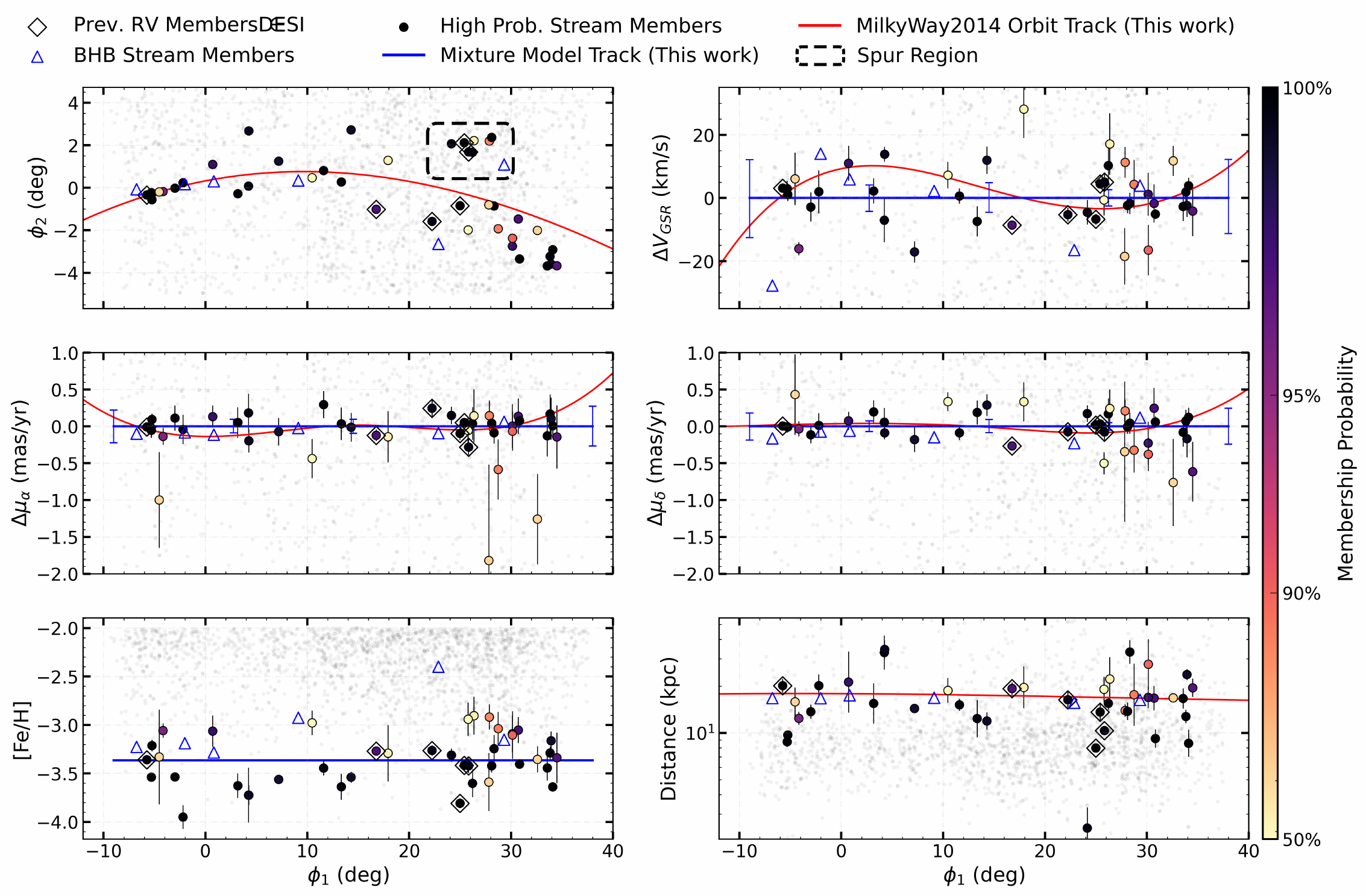}
    \caption{41 high-probable $(P_{\text{s}} > 50\% , \text{see colourbar})$ member stars in the C-19 stellar stream plus 6 blue horizontal branch members (blue triangle markers) obtained from our box-cuts about the spline tracks. The light grey points are `low probability' $(P_{\text{s}} > 50\%)$ stars. We plot stars in residual space for  $\Delta \mu_\alpha, \Delta\mu_\delta, \text{and } \Delta v_\text{GSR}$. We show the fitted spline tracks from the mixture model in blue for the modelled parameters. The blue error bars along the spline track show the node locations for each spline, and the uncertainties are obtained from the 65\% posterior credible intervals. In all panels but metallicity, we show the orbit track (see \S \ref{sec:orbit}) for a C-19 progenitor within the \texttt{galpy MilkyWay2014} potential \citep{bovy_galpy_2015} in red. In the top-left panel, we outline the spur region in black; this same region is shown in Figures \ref{fig:vdisp} and \ref{fig:box_sky}. Figure \ref{fig:6_app} in Appendix \ref{app:mems} shows this figure in physical space, rather than as residuals with respect to the spline tracks.}
    \label{fig:memprob}
\end{figure*}

The distribution of assigned membership probabilities is shown in Figure \ref{fig:memhist}, displaying a very clear distinction between probabilities given to member stars and background stars. In Figure \ref{fig:memprob}, we show our identified member stars of C-19, coloured by their membership probability. We find 41 high-probable $\left(P_{\in\text{ stream}} > 0.5 \right)$ MS+RGB stars. Among the 23 spectroscopically confirmed C-19 members reported by \citet{yuan_pristine_2022, yuan_pristine_2025}, 4 have DESI DR2 spectroscopy. Including their ‘probable’ members increases this to 6 stars observed in DESI DR2, all of which have membership probabilities above 95\% from our mixture model.

Our member stars span $\sim 45^{\circ}$ in $\phi_1$, limited by the coverage of the MWS (see Figure \ref{fig:footprint}).  The spatial distribution of member stars we see for the C-19 stream agrees with the distribution of members determined by \citet{yuan_pristine_2025}, which extends past the DESI footprint to span $\sim 100^{\circ}$. We measure a mean metallicity for C-19 of $\text{[Fe/H] =}{-3.36_{-0.10}^{+0.12}}$ dex, reaffirming its extremely metal-poor nature. Our ability to place constraints on the observed metallicity dispersion is discussed further in \S  \ref{sec:metal}.

\begin{figure}
    \centering
    
    \includegraphics[width=1\linewidth]{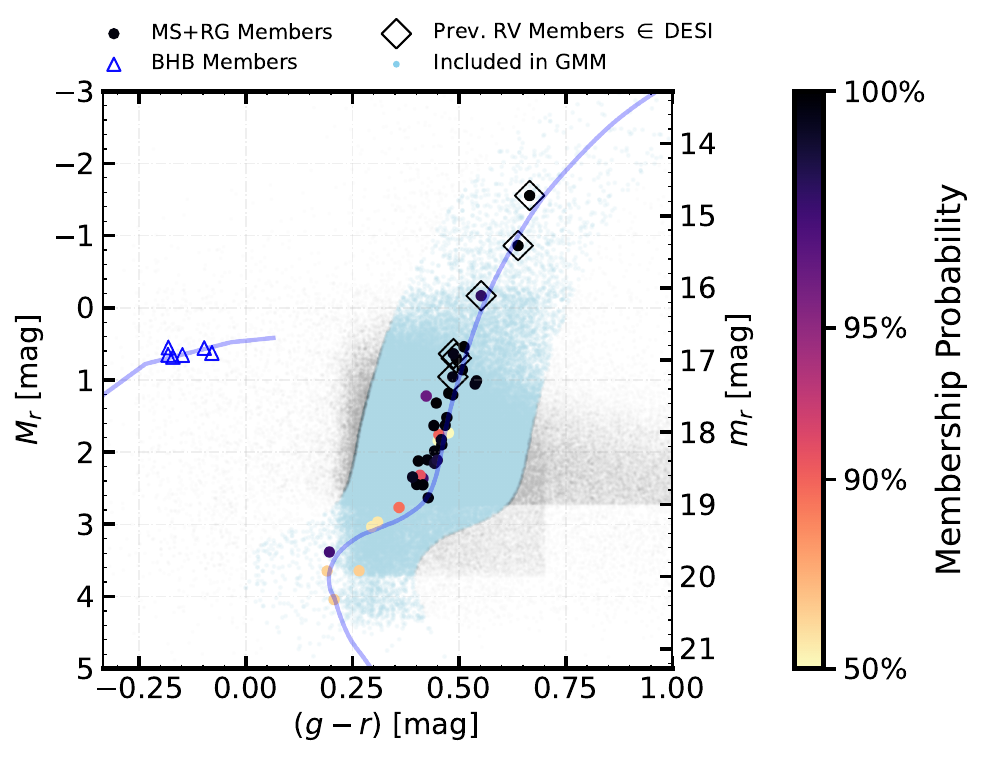}

    \caption{Absolute and apparent colour-magnitude diagram at the distance of C-19 ($18$ kpc); the apparent magnitude scale shown on the right $y$-axis. The 41 high-probable MS+RGB members are shown with their inferred stream membership probability (see colourbar to the right). We highlight the 6 previously confirmed members radial velocity members with diamond markers (\citey{yuan_pristine_2022};  \citey{yuan_pristine_2025}), which are all bright giants past the main sequence. We show our \citet{dotter_mesa_2016} MS+RGB isochrone at [Fe/H] = $-2.49$ and an age of 13.5 Gyr as the right blue curve; the DESI stars that pass our 0.20 \textit{(g-r)} magnitude tolerance cut and make it into the mixture model are coloured light blue.  The 6 Horizontal branch members from \S \ref{sec:box-cuts} are shown as blue triangle markers alongside the empirical horizontal branch of M 92 from the \citet{harris_catalog_1996} catalogue in blue.}
    \label{fig:cmds}
    
\end{figure}

In the bottom right panel of Figure \ref{fig:memprob}, we show the heliocentric distances of high membership probability stars as described in \S \ref{sec:data}. Although we do not include any distance information in our mixture model, we use it to check the results of our method; real stream members would follow a coherent distance track. We see a continuous track with no significant distance gradient in our high-probable C-19 members, in line with previous results for this segment of the stream \citep{martin_stellar_2022, yuan_pristine_2022}. Moreover, the member stars are all found at larger heliocentric distances than the background population, consistent with them belonging to a distinct population. We also plot our member stars in colour-magnitude space in Figure \ref{fig:cmds}. Even without including colour in our likelihood function, the member stars cluster with little scatter about the metal-poor isochrone. We highlight the 6 member stars from \citet{yuan_pristine_2025} with diamond markers; the majority of the stars we identify are much fainter, approaching the main sequence turn-off. 

We note a novel `spur' feature in the spatial projection of C-19 at $\sim \phi_1 = 25^\circ$, shown in the top left panel of Figure \ref{fig:memprob}. The spur feature is outlined by a dotted black box for clarity. This structure is similar to features previously identified in the Grillmair Dionatos (GD-1) \citep{grillmair_detection_2006,price-whelan_off_2018,bonaca_spur_2019} and the ATLAS Aliqa-Uma (AAU, \citealt{li_broken_2021}) streams. Such structures in streams point towards past interactions with other substructure within the MW halo, and have been used as a constraint on the dark matter subhalo mass function. We discuss the spur further in \S \ref{sec:spur}.

\begin{table*}
\centering
\caption{Spline tracks fit for C-19. 5 equally spaced knots are used, with cubic splines interpolated between them for $E_{v_\text{GSR}}$, $E_{\mu_\alpha}$, and $E_{\mu_\delta}$. In our base model, the stream fraction $f_s$ and kinematic dispersions are held constant across the stream's extent; these are allowed to vary with no change in our stream membership.} 
\begin{tabular*}{\linewidth}{@{\extracolsep{\fill}} l r r r r c c c c}\label{tab:spline}

& & \multicolumn{3}{c}{Spline nodes} & & \multicolumn{3}{c}{Constant} \\
\cline{3-5} \cline{7-9}

$i$ &
$\phi_{1,i}$ (deg) &
$E_{v_{\mathrm{GSR}},i}$ (km\,s$^{-1}$) &
$E_{\mu_{\alpha,i}}$ (mas\,yr$^{-1}$) &
$E_{\mu_{\delta,i}}$ (mas\,yr$^{-1}$) &
&
$f_s$ (\%) &
$\sigma_{v_{\mathrm{GSR}}}$ (km\,s$^{-1}$) &
$^\ddagger\sigma_{\mu_\alpha}=\sigma_{\mu_\delta}$ (mas\,yr$^{-1}$) \\
\hline

1 & $-9.0$   & $26.4_{-11.8}^{+12.2}$  & $1.18_{-0.21}^{+0.26}$   & $-2.5_{-0.2}^{+0.2}$ &
& \multirow[c]{5}{*}{\centering $2.0_{-0.3}^{+0.3}$} &
  \multirow[c]{5}{*}{\centering $7.8_{-1.3}^{+1.5}$} &
  \multirow[c]{5}{*}{\centering $0.10$} \\
  
2 & $2.75$   & $-9.8_{-4.2}^{+4.1}$    & $1.25_{-0.09}^{+0.09}$   & $-3.1_{-0.1}^{+0.1}$ &
& & & \\

3 & $14.5$   & $-15.6_{-4.3}^{+4.4}$   & $0.92_{-0.09}^{+0.09}$   & $-3.6_{-0.1}^{+0.1}$ &
& & & \\

4 & $26.25$  & $-18.2_{-2.7}^{+2.6}$   & $0.74_{-0.05}^{+0.05}$   & $-4.1_{-0.1}^{+0.1}$ &
& & & \\

5 & $38.0$   & $-33.9_{-10.6}^{+11.2}$ & $-0.14_{-0.28}^{+0.29}$  & $-5.2_{-0.2}^{+0.2}$ &
& & & \\
\hline

\end{tabular*}
\raggedright
\footnotesize
$^\ddagger$ This value was fixed to be equivalent to a dispersion of $\sim 8 \text{ km s}^{-1}$ at a distance of $18$ kpc.
\end{table*}

\subsection{Model Validation}\label{sec:verify}

We remark that we adopt a simplified approach in our employment of a two-component Gaussian mixture model; the future objective is to apply our approach to many streams in DESI while needing few changes from stream-to-stream. However, we must justify the assumptions implicit in our generalization. In our mixture model we posit three key assumptions. For one, we assume a constant stream fraction along the length of C-19. While this is a reasonable approximation, the presence of over- and under-densities within any given stream, the trend that the member density of a stream declines as you move further from the progenitor's original location, and background density changes, all suggest that the stream fraction could vary along its extent. Secondly, the mixture model we use holds constant the dispersion in $v_\text{GSR}$; this constraint may overestimate the overall heating of the stream if only a portion of it has been significantly perturbed. Third, we assume that the mean and dispersion of all of the background parameters are constant across the modelled area, which may not be the case along some lines of sight. 

As noted in \S \ref{sec:mcmc}, we see no change in the membership results when we remove the first two assumptions. We further assess the impact of this choice by drawing posterior-predictive realizations from our fitted model. Specifically, we generate mock data from our best-fit models with observational uncertainties and compare their distributions to those of the real data (see Figures \ref{fig:ppc_all}, \ref{fig:ppc_qq}). These diagnostic steps show that our model captures the overall distribution of the background and stream stars---we go into greater detail on this in Appendix \ref{app:ppc}. However, we do not run a fit in which the background is allowed to vary as a function of $\phi_1$, and the posterior realization check is not well suited to assess if a varying background is a better suited choice. We therefore perform stream selection using a method that does not rely on any explicit background model in \S \ref{sec:box-cuts}.

\subsubsection{Box-cuts} \label{sec:box-cuts}

\begin{figure}
    \centering
        \includegraphics[width=0.95\linewidth]{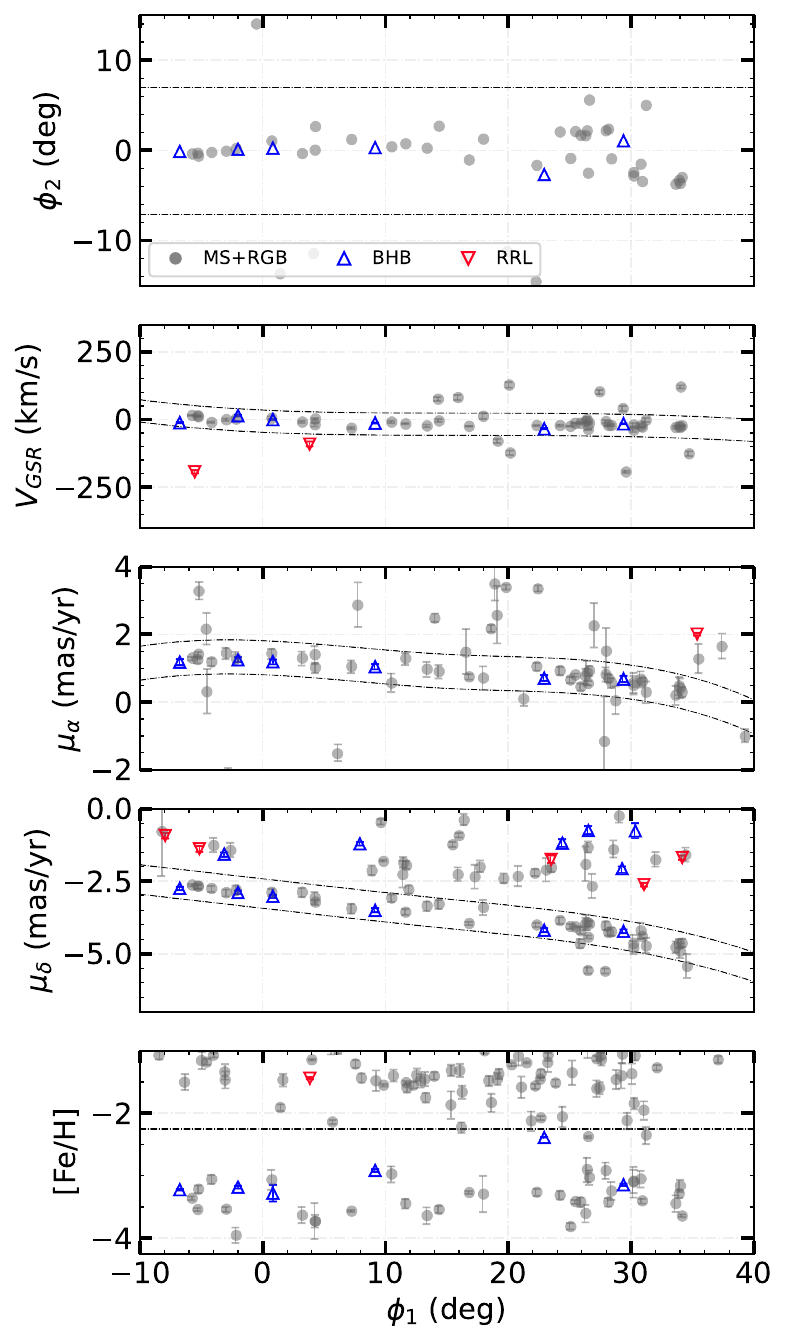}
    
    \caption{Track selection of likely C-19 members. For each panel, showing one phase-space coordinate as a function of $\phi_1$, we plot the stars remaining after all other selection cuts are shown in each panel (i.e. the $v_\text{GSR}$ panel shows stars selected with $\phi_2$, metallicity, and proper motions). Grey circles represent MS+RGB stars, blue upper triangles represent BHB stars in \citet{bystrom_exploring_2024}, red lower triangles denote RRL stars in \citet{medina_desi_2025}. The dashed lines show the tolerance about the spline track fit in \S \ref{sec:mcmc} used to select likely member stars. In each panel, the stars trace a coherent sequence along the spline track even before the final box cut is applied. The final selection of box-cut members includes only stars that survive selection across all panels.}
    \label{fig:boxcut}
\end{figure}

We assess the impact of assuming a constant background by constructing a set of likely stream members without explicit background modelling, instead applying tight cuts around the modelled stream track and the best-fit isochrone. Our cuts are as follows: $\pm7.5^\circ$ degrees about $\phi_2 = 0$, $\pm 40 \text{ km s}^{-1}$ about the velocity track,  $\pm 0.5 \text{ mas yr}^{-1}$ about each proper motion track, an upper bound of $-2.25$ dex in [Fe/H], $\pm 10^{\circ} \text{ in } \phi_2$, and $\delta(g-r) = 0.075$ about the same MS+RGB stellar isochrone used in \S \ref{sec:trunc}. We show the tolerance about the spline tracks for $v_\text{GSR}$, $\mu_\alpha$, $\mu_\delta$, and [Fe/H] in Figure \ref{fig:boxcut}. Additionally, we allow horizontal branch stars to be considered in the box-cut approach, placing the horizontal branch of the M 92 GC from the \citet{harris_catalog_1996} at the C-19 distance of 18 kpc.

Each panel in Figure \ref{fig:boxcut} plots the stars remaining after all of the \textit{other} cuts have been made; the purpose of this is to demonstrate that, in each of the modelled spaces ($v_\text{GSR}, \mu_\alpha, \mu_\delta, [\text{Fe/H}]$), a coherent structure of stars follow the track without applying the final selection cut (shown as dashed lines).  For example, the second panel of Figure \ref{fig:boxcut} plots $v_\text{GSR}$ against $\phi_1$ while applying cuts in $\phi_2$, proper motion, and metallicity. Despite no cuts made in line-of-sight velocity, we see the majority of the remaining stars cluster about the spline track.

We cross-match our resulting box-cut members with the DR2 versions of the BHB \citep{bystrom_exploring_2024} and RR Lyrae (RRL, \citey{medina_desi_2025}) catalogues. In all, we find 38 MS+RGB likely members using the box-cut approach, 35 of which were also found by the mixture model. We also see 6 horizontal branch members, all of which are BHBs. These members are provided in Table \ref{tab:members2}. The top panel of Figure \ref{fig:boxcut} presents the spatial selection for the box-cut member, within which the spur substructure is retrieved.

\subsection{Velocity Dispersion} \label{sec:vdisp}

We find a velocity dispersion of $7.8_{-1.3}^{+1.5} \text{ km s}^{-1}$ for the stream population in our mixture model. The line-of-sight velocity of C-19 has been a point of discussion, as past measurements ($\sim 7$ km s$^{-1}$, \citealt{martin_stellar_2022}; $11.1_{-1.6}^{+1.9}$ km s$^{-1}$, \citealt{yuan_pristine_2025}) have found the stream to be dynamically hot compared to known GC streams ($\sim2-5 \text{ km s}^{-1}$, \citealt{li_s_2022}).

\begin{figure*}[!th]
    \centering
    \includegraphics[width=0.85\linewidth]{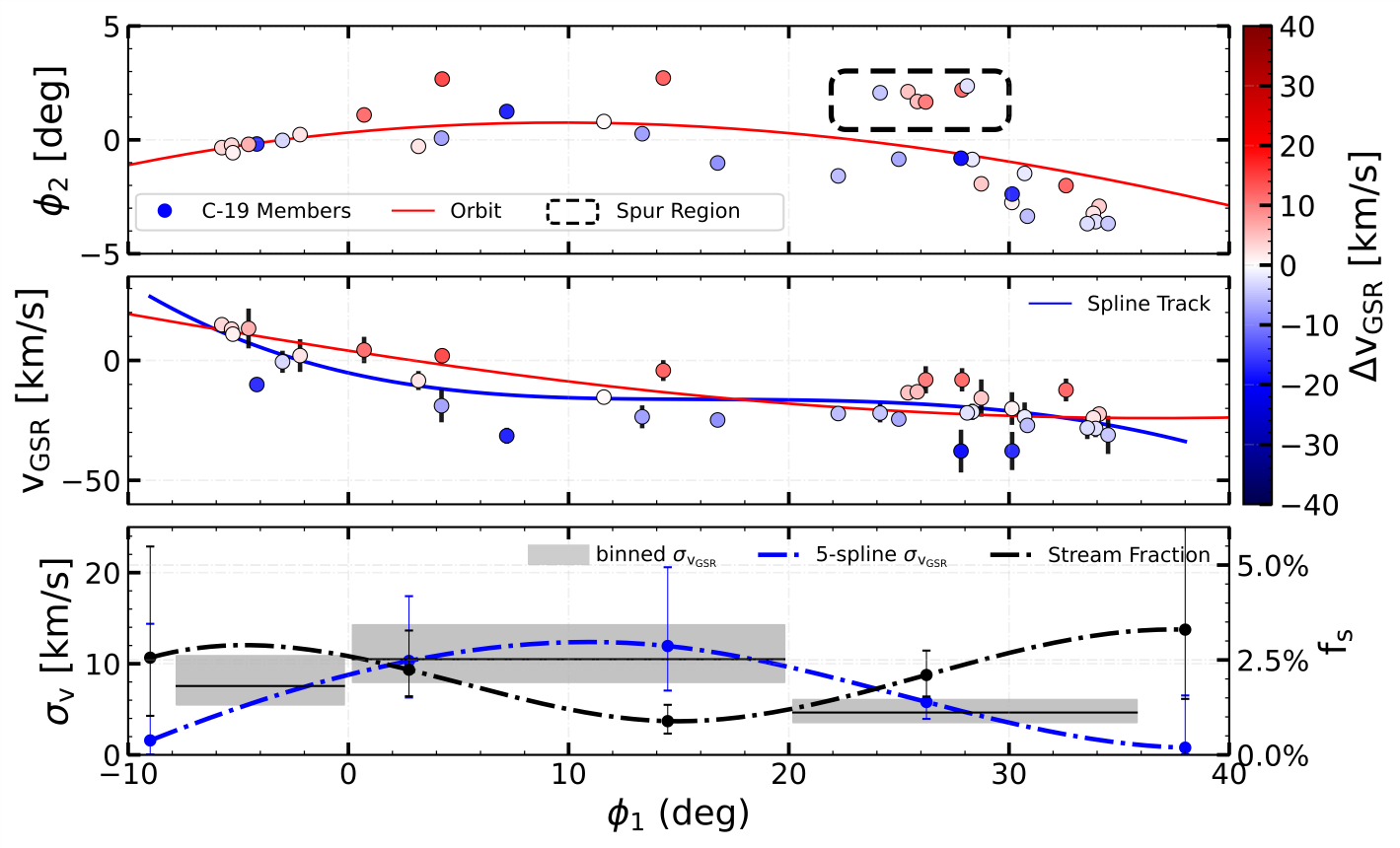}
    \caption{\textit{(top)} On-sky distribution of C-19 members with $P_{\in\text{ stream} } > 75\%$, coloured by distance from spline $v_\text{GSR}$ track. The spur region outlined in black is the same as those shown in Figures \ref{fig:memprob} and \ref{fig:box_sky}. \textit{(middle)} $v_\text{GSR}$ space for C-19. \textit{(bottom)} $\sigma_{v_\text{GSR}}$ calculated in three bins along the stream's position in $\phi_1$. The mean and spread in $\sigma_{v_\text{GSR}}$ are shown using black horizontal lines with shaded grey regions. Overlayed are the spline tracks of $\sigma_{v_\text{GSR}}$ (blue dot dash) and stream fraction (black dot dash) once we allow it to vary in our mixture model; these correspond to the y-axis ticks on the left and right, respectively. The median posterior value of our non-varying $\sigma_{v_\text{GSR}}$ is $7.8^{+1.5}_{-1.3} \text{ km s}^{-1}$. We show the orbit track (see \S \ref{sec:orbit}) for a C-19 progenitor within the \texttt{galpy MilkyWay2014} potential \citep{bovy_galpy_2015} in red in the first two panels.}
    \label{fig:vdisp}
\end{figure*}

We take care to also study how the velocity dispersion varies along the stream's track, defining $\Delta v_\text{GSR}$ as the difference between each star's $v_\text{GSR}$ and the stream track in velocity space. This is shown in Figure \ref{fig:vdisp}, where we divide C-19 into three tentative sections: the `thin' end, $\phi_1 \in (-10^\circ, 0^\circ)$,  the sparse middle, $\phi_1 \in (0^\circ, 20^\circ)$, and the on-spur, $\phi_1 \in( 20^\circ, 35^\circ)$. We model the velocity distribution in each section as a Gaussian, whose width is the intrinsic dispersion, and sample the posterior using the \texttt{emcee} ensemble sampler \citep{foreman-mackey_emcee_2013}. We use 50 walkers for 1100 steps and discard the first 100 steps as burn-in, performing the fit on C-19 members with $P_{\in \text{ stream}} > 75\%$. We trim for veryhighly probable members because we do not include membership probability in this model. From left-to-right in $\phi_1$, we find velocity dispersions of $7.7_{-3.3}^{+2.8}$, $10.6_{-2.5}^{+3.9}$, and $3.9_{-1.0}^{+1.3}$ km s$^{-1}$ (with 7, 11, and 23 stars in each segment respectively).

In the bottom panel of Figure \ref{fig:vdisp},  we overplot the spline tracks for both $\sigma_{v_\text{GSR}}$ and stream fraction, which we obtain by increasing the number of spline points for both from 1 to 5 (the 35-parameter model from \S \ref{sec:mcmc}). For $\sigma_{v_\text{GSR}}$, this is a good check of our segmented velocity dispersion measurements, as now we are not making abrupt bounds, but rather allowing the velocity dispersion of each point to be informed by the entire stream.  Qualitatively, both the segmented and spline methods agree on the trend of velocity dispersion along $\phi_1$. Interestingly, we find that the stream fraction (black dot dash, bottom panel of Figure \ref{fig:vdisp}) is anti-correlated with the velocity dispersion (blue dot dash, bottom panel of Figure \ref{fig:vdisp}), with higher velocity dispersions corresponding to lower stream fractions. This could indicate that the segment at $\phi_1 \in (0^\circ, 20^\circ)$ experienced a past perturbation, producing a lower-density region or gap with hotter kinematics. Alternatively, the low density of member stars in this area may lead to a higher level of contamination from the background, artificially inflating the measured velocity dispersion. Deeper observations will be required to distinguish between these two scenarios.

\subsection{`Spur' Feature}\label{sec:spur}

    Spur-like perturbations have been identified in a small number of stellar streams, most notably GD-1 \citep{price-whelan_off_2018, bonaca_spur_2019} and AAU \citep{li_broken_2021}. C-19's spur is offset from the main body by $\sim 1^\circ$, corresponding to a physical offset of $\sim 300$ pc at a distance of 18 kpc. This separation is approximately three times that of the GD-1 spur, and is a similar offset to that of the kink in AAU ($\sim 400$ pc). The length of the spur is $\sim 900$ pc ($\sim 3^\circ$), comparable to the $\sim 1,300$ pc extent of the GD-1 spur. If this is indeed a newly identified spur, its physical properties are broadly consistent with those observed in previously identified spurs.
    
    \citet{li_broken_2021} showed that the AAU stream members exhibit strong correlations between their on-sky displacement from the stream track and their line-of-sight velocities. For C-19 stream members, colouring the stars by their distance from the $v_\text{GSR}$ spline in the top two panels of Figure \ref{fig:vdisp}, we see that spur members (outlined by the dotted black box) tend to lie above the $v_\text{GSR}$ track, while stars at similar $\phi_1$ but with $\phi_2 < 0$ preferentially fall below it. We measure the kinematic offset of the C-19 spur member stars to be about $10$ km s$^{-1}$, which is of the same order of the $>20$ km s$^{-1}$ offset seen in AAU. Despite the fact that the GD-1 spur does not show any such velocity offset, the agreement with AAU's kinematics bolsters the argument for the C-19 spur being a real feature. 


\section{Orbit Modelling} \label{sec:orbit}

Modelling the orbit of halo stars has been a popular method for both searching for undiscovered streams and finding new members for known streams (\citealt{ibata_phlegethon_2018, ibata_charting_2021,li_broken_2021,martin_stellar_2022,ibata_charting_2023, awad_s5_2025,yuan_pristine_2025, aganze_cocytos_2025}, for example). Furthermore, modelling the orbit of known streams lends insight into the shape, mass, and clumpiness of the Galactic halo \citep{bovy_dynamical_2014, bovy_shape_2016, nibauer_slant_2024, nibauer_textttstreamsculptor_2024}. Thus far, we avoid using orbit modelling in our member selection to keep our stream selection unbiased by an assumed underlying Galactic potential. Instead, we model the stream's orbit after member selection and use it to verify our results are in line with the \citet{yuan_pristine_2025} members that are beyond the MWS footprint. 


\begin{table}
\centering
\caption{Pseudo-progenitor phase-space coordinates and orbital properties, modelled in \texttt{MWPotential2014}. We fixed the progenitor's $\alpha$ value to be the centre of our observed stream extent.}
\begin{tabular*}{0.85\linewidth}{@{\extracolsep{\fill}} l c}
\hline
Property & Value \\
\hline
$\alpha$ (deg) & $351.9$ \\
$\delta$ (deg) & $2.8$ \\
$\mu_{\alpha}$ (mas\,yr$^{-1}$) & $0.8$ \\
$\mu_{\delta}$ (mas\,yr$^{-1}$) & $-4.0$ \\
$v_{\mathrm{los}}$ (km\,s$^{-1}$) & $-158.0$ \\
Distance (kpc) & $17.1$ \\
Pericenter (kpc) & $10.2$ \\
Apocenter (kpc) & $21.4$ \\
Orbital period (Myr) & $370$ \\
Eccentricity & $0.36$\\
\hline
\end{tabular*}
\label{tab:orbit}
\end{table}

We fit an orbit for the C-19 progenitor using 41 high-probability stream members, each of which has full five-dimensional phase-space information ($\alpha$, $\delta$, $\mu_\alpha$, $\mu_\delta$, $v_\text{GSR}$). We also have distance values from \texttt{rvsdistnn} but do not include them in the fit due to their large uncertainties. Although each individual star follows its own distinct orbit, the ensemble of stars approximately traces the orbit of the now-dissolved progenitor. For the purpose of orbit modelling, we model the orbit of a pseudo-progenitor and fix its right ascension to $\alpha = 351.9^\circ$, corresponding to the midpoint of the stream in right ascension or a $\phi_1$ of $21.5^\circ$. The remaining five phase-space coordinates of the progenitor, $(\delta$, distance, $\mu_\alpha$, $\mu_\delta$, and $v_\text{GSR})$, are treated as free parameters. Our initial guess for these parameters is taken to be the average of the member stars closest to the stream midpoint. For a given set of progenitor coordinates in six-dimensional phase space, we integrate the corresponding orbit using \texttt{galpy} in the \texttt{MilkyWay2014} potential \cite{bovy_galpy_2015}. The resulting orbit is expressed as a function of $\phi_1$. We evaluate the likelihood of the model by assuming that, at each star’s $\phi_1$, the observed values of $\phi_2$, $\mu_\alpha$, $\mu_\delta$, and $v_\text{GSR}$ are independently drawn from Gaussian distributions centered on the model orbit, with standard deviations given by the observational uncertainties. We determine the best-fitting progenitor parameters by maximizing this likelihood, and report the resulting fit in Table \ref{tab:orbit}.

We show the derived orbit track over $\sim$ 200 Myr (integrating both forwards and backwards from the present day) in Mollweide projection (Figure \ref{fig:footprint}) and in 6D space (Figure \ref{fig:memprob}). We see strong agreement between the spline tracks and the orbit track in the kinematic panels $v_\text{GSR}, \mu_\alpha,$ and $\mu_\delta$ of Figure \ref{fig:memprob}.  In the top panel, we see that the majority of the stream's members with $\phi_1 > 20^\circ$ fall below the orbit track in $\phi_2$, the exception is the spur at $\phi_1 \sim 25^\circ$, making the substructure more distinct. We do not use distance in our model likelihood yet observe an agreement between the orbit distance track and the distance of our member stars. The distance track shows a very shallow, if any, distance gradient over this segment of the stream. The orbit fitting also gives an estimate for the width of the stream. Because we are fitting for an on-sky track, we find an intrinsic dispersion in $\Delta\phi_2$ of $1.50^\circ$. At a distance of $18$ kpc, this corresponds to a physical width of $\sim$ 450 pc, which is wider than normal GC streams \citep{li_s_2022}. 

\begin{figure}[h]
    \centering
    \includegraphics[width=1\linewidth]{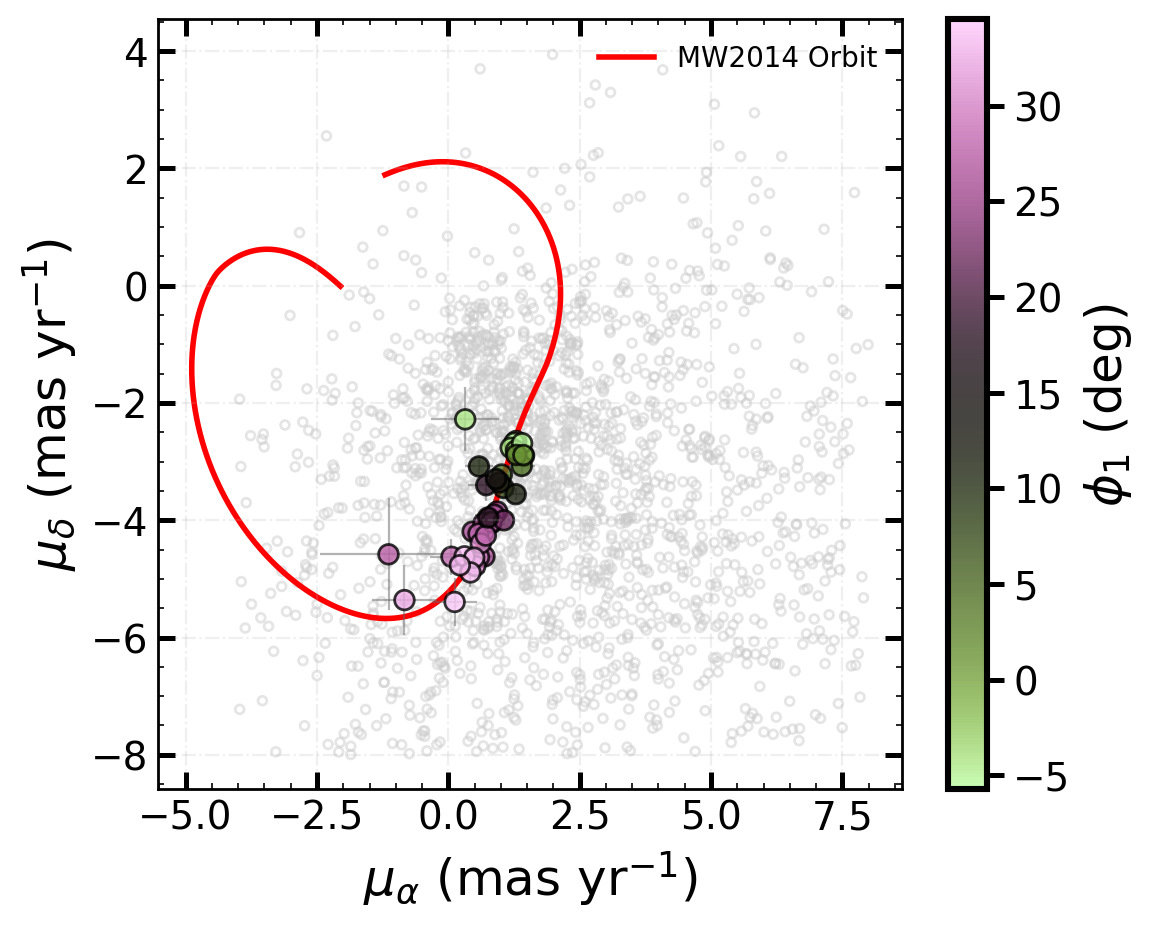}
    \caption{Proper motion distribution of the high probability member stars from Figure \ref{fig:memprob}. The colours of the points mark the position of the members in $\phi_1$ along the stream's extent. The background shows the distribution of all MWS DR2 members that survived our cuts in \S \ref{sec:trunc}.}
    \label{fig:proper_motions}
\end{figure}

We can view the stream members and our fitted orbit in proper motion space, shown in Figure \ref{fig:proper_motions}.  The stream members follow the orbit in proper motion space with a consistent gradient in $\phi_1$. The background stars shown here are those stars that survive the data truncation of \S  \ref{sec:trunc}; notably, Figure \ref{fig:proper_motions} demonstrates that if our membership selection relied on making a cut in proper motion space along our orbit track, with no spectroscopic information, we would include many contaminants. 
We see this realized by the fact that, of the six \texttt{STREAMFINDER} \citep{ibata_charting_2023} members of C-19, half are immediately ruled out of membership from their iron content (Figure \ref{fig:trunc}, bottom panel), and two-thirds stray from the stream's track in $v_\text{GSR}$ (Figure \ref{fig:trunc}, top panel); only 2 of the \texttt{STREAMFINDER} members in the MWS DR2 are confirmed by our mixture model approach.

\section{Discussions}\label{sec:discussion}

\subsection{C-19 in \textit{Gaia}+DECaLS}\label{sec:gaiadecals}
The MWS DR2 suffers completeness variations, as we discuss further in \S \ref{sec:complete}. For this reason, we perform an independent investigation of C-19 as viewed in the \textit{Gaia} and DECaLS surveys (which probe as deep as $G \sim 20.5$ and $r \sim 23.5$ mag respectively while maintaining uniform depth \citealt{gaia_collaboration_gaia_2016, dey_overview_2019}). We cross-match each DECaLS star with a 1 arcsecond tolerance using a spherical distance match to the nearest \textit{Gaia} DR3 source. We remove duplicates in the combined dataset by keeping instances with the smallest difference between the \textit{Gaia} $G$ and DECaLS $r$ magnitudes. Next, we apply cuts informed by our modelled stream track to the combined \textit{Gaia}+DECaLS dataset. We note that the selection criteria are intentionally stringent to isolate the purest possible C-19 sample from the Gaia+DECaLS data set.

\begin{figure}[!t]
    \includegraphics[width=0.95\linewidth]{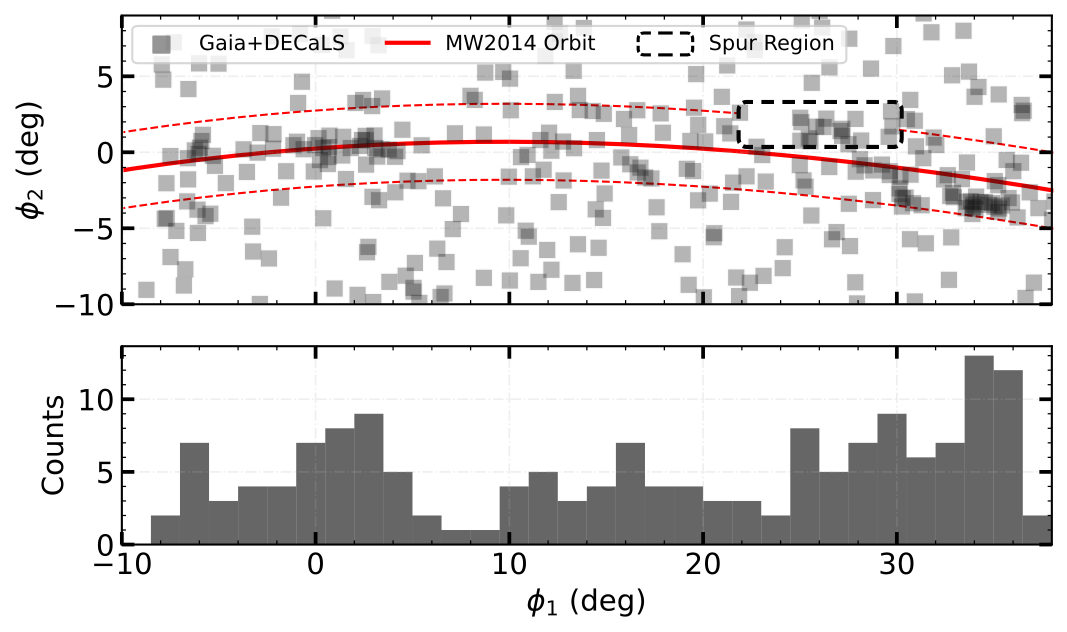}
    \caption{\textit{(top)} Box-cuts about spline track applied to \textit{Gaia}+DECaLS data down to $20.5$ r-band magnitude. Orbit model track shown in red. The spur region outlined in black is the same as those shown in Figures \ref{fig:memprob} and \ref{fig:vdisp}. \textit{(bottom)} Binned counts of stars within $2.5^\circ$ of the orbit track, where each bin is $1.5^\circ$ in $\phi_1$. This view highlights the density variations apparent across the stream from our mixture model results. The highest peak ($\phi_1 \sim 34^\circ$) coincides with an area of low completeness in the DESI MWS DR2 dataset; the densest clump almost entirely falls within a patch of the sky not yet observed.}
    \label{fig:box_sky}
\end{figure}

The top panel of Figure \ref{fig:box_sky} shows the remaining stars after cuts in proper motion of $\pm0.3 \text{ mas yr}^{-1}$ relative to the proper motion stream tracks, $\pm 0.05  \ \delta(g-r)$ colour index with respect to our selected isochrone, a parallax cut as Equation \ref{eq:plx}, a magnitude cut between $r \in [16, 20.5]$ mag, and a photometric metallicity cut qualitatively made with a selection in $r-z$ vs $g-r$ space to select metal-poor stars (as done in \citey{li_southern_2019}). With these cuts applied, we observe similar substructure within C-19 that we do in DESI, particularly that of the sparse middle region between clumps centred on $\phi_1 \sim 1^\circ$ and $\phi_1 \sim 34^\circ$. In the bottom panel of Figure \ref{fig:box_sky}, we show the count of \textit{Gaia}+DECaLS stars within $2.5^\circ$ of our modelled orbit in $1.5^\circ \ \phi_1$ bins. The $\phi_1$ variations in star count follow the same trend as the stream fraction spline when we allow it to vary (Figure \ref{fig:vdisp}). 

Finally, in Figure \ref{fig:box_sky} we overlay the same spur region identified by eye in Figure \ref{fig:memprob}. Within this region, there is a tentative over-density in the \textit{Gaia}+DECaLS selection. However, the structure is not significant enough to conclusively determine the existence of a spur in C-19. Deeper spectroscopic follow up is needed for confirmation.

\subsection{Completeness in DESI MWS DR2}\label{sec:complete}

\begin{figure}[!t]
    \centering
    \includegraphics[width=1\linewidth]{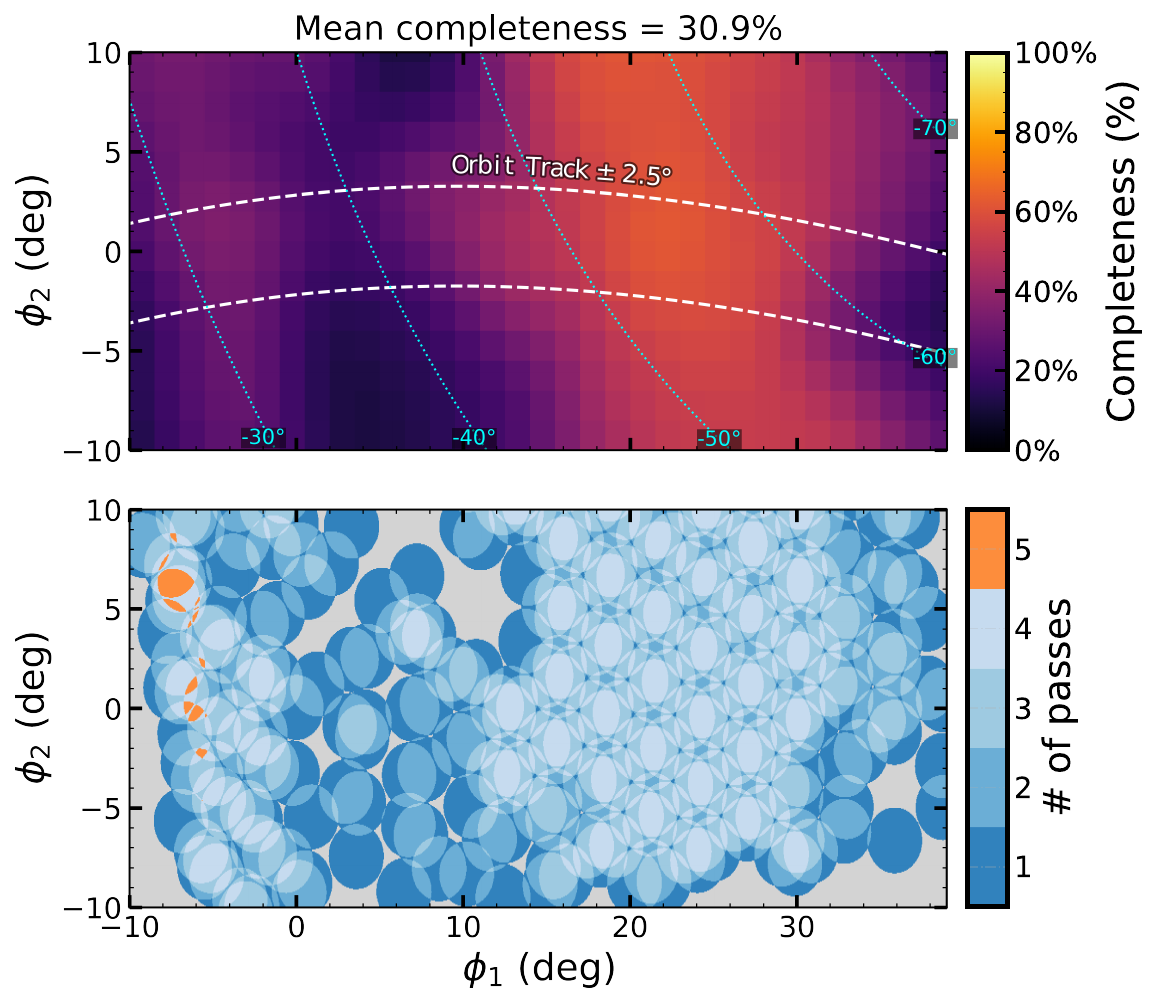}
    \caption{\textit{(top)} Completeness of the MWS DR2, measured relative to Gaia+DECaLS for $r \in [16.0, 19.0]$. The map is binned in $1.5^\circ$ pixels and smoothed with a Gaussian filter of width 1.5 bins. We show lines of equal Galactic Latitude in cyan, and the $\pm 2.5^\circ$ padding about the orbit track from the top panel of Figure \ref{fig:box_sky}. \textit{(bottom)} Number of passes for each part of the sky by the DESI MWS after three years of observations.}
    \label{fig:hist}
\end{figure}

The DESI MWS target selection was accomplished using DECaLS, therefore, we can plot the spatial trend in survey completeness using the ratio between the number of stars as C-19 targets in \textit{Gaia}+DECaLS to the count observed in MWS DR2 within the same magnitude range. We reduce the \textit{Gaia}+DECaLS and MWS data with proper motion cuts ($\pm 1.5$ mas yr$^{-1}$) relative to the fit stream track, the parallax cut from Equation \ref{eq:plx}, and a $0.15  \ g-r$ colour index cut with respect to the selected isochrone. We visualize the completeness along the sky using a 2D histogram in the top panel of Figure \ref{fig:hist} for $r_\text{mag} \in [16, 19]$, binned in $1.5^\circ$ pixels. The magnitude cut restricts both data sets only to the magnitude range the DESI MWS was designed to target. To mitigate spurious fluctuations in completeness in Figure \ref{fig:hist}, we apply a Gaussian filter with a kernel width of 1.5 bins to highlight coherent spatial trends. 

We find that the MWS DR2 has a completeness of $30.5 \%$ in the C-19 region, with the area most complete ($\phi_1 \in \sim [20^\circ, 30^\circ]$) overlapping the on-spur and sparse-middle regions. The bottom panel of Figure \ref{fig:hist} shows the number of passes the MWS DR2 has completed for the region from the DESI main-bright program. As expected, the similarities between the number of passes and high completeness as measured against \textit{Gaia}+DECaLS are apparent; the spur feature is relatively well sampled in this regard. It is notable that the areas with the fewest passes/lowest completeness coincide with the largest overdensities in the C-19 structure around $\phi_1=0^\circ$ and $\phi_1=34^\circ$. For example, the densest concentration in the top panel of Figure \ref{fig:box_sky} aligns with the area of reduced DR2 completeness visible in the bottom-right region of both panels in Figure \ref{fig:hist} (around $\phi_1\sim 34^\circ)$. We anticipate a substantial increase in membership in these regions once the MWS is complete, with the total number of confirmed C-19 member stars expected to grow by a factor of 2-3 by the end of DESI observations.

\subsection{Progenitor Properties}\label{sec:prog}

\subsubsection{Metallicity Constraint}\label{sec:metal}


Metallicity dispersion is a key diagnostic of stream progenitor origin. GCs exhibit very small intrinsic spreads in [Fe/H], typically $\lesssim 0.1$ dex, whereas dwarf galaxies show substantially broader metallicity distributions, commonly $\gtrsim 0.25$ dex \citep{li_s_2022}. Our model yields a best-fit intrinsic metallicity dispersion for C-19 of $\sigma_{\text{[Fe/H]}} = 0.23^{+0.05}_{-0.04}$ dex, which is larger than the previous upper limit of $< 0.18$ dex (95\% confidence) reported by \citet{martin_pristine_2022}. While this value lies near the lower end of the metallicity dispersions typically observed in dwarf galaxies, we caution against interpreting it as strong evidence for a dwarf-galaxy progenitor.

In particular, the reported [Fe/H] uncertainties reflect formal measurement errors only and do not account for systematic effects that become significant in the extremely metal-poor regime. Such metallicity-dependent systematics, including limitations in spectral modelling and line diagnostics, can introduce star-to-star abundance biases that artificially inflate the inferred intrinsic dispersion, even for chemically homogeneous systems. This effect is clearly demonstrated in the DESI DR1 Stellar Catalogue, where several well-studied, metal-poor globular clusters and streams (e.g. M 92, M 15, and NGC 5053, all with [Fe/H] $< -2.0$ dex) exhibit metallicity dispersions that are substantially larger than literature values (see Table 7 and Appendix D of \citealt{KoposovSE2025}).

We therefore interpret our measured dispersion of $\sigma_{\text{[Fe/H]}} = 0.23^{+0.05}_{-0.04}$ dex as an upper bound on the true intrinsic metallicity spread of C-19. Given the likely contribution of systematic abundance biases, this measurement alone is insufficient to discriminate conclusively between a globular cluster or dwarf galaxy progenitor.

\subsubsection{Mass Estimation}\label{sec:mass}
We estimate the mass of the stream within the observed range $\phi_1 \in [-10^\circ, 40^\circ]$. Our tracks are well-suited for estimating the mass, as they are informed by high-probability stream members much fainter than previous results (see Figure \ref{fig:cmds}). First, we quantify the overdensity in stars due to C-19's presence in the \textit{Gaia}+DECaLS field; we trim the \textit{Gaia}+DECaLS dataset to $\pm 0.4 \text{ mas yr}^{-1}$ in $\mu_\alpha \text{ and } \mu_\delta$, $\pm 7^\circ$ in $\phi_2$ about our stream track and $\pm 0.05 \ (g-r)$ about our selected isochrone, with the rest of the cuts being the same as \S \ref{sec:complete}. We then fit a simple two-component mixture model to the distribution of stars in $\Delta\phi_2$ around the orbit track. In this model, the background population is taken to be uniformly distributed in $\Delta\phi_2$, while the stream population is described by a normal distribution centred on the orbit track. This gives $\sim$ 200 stars belonging to C-19 between $r \in [16, 20.5]$ mag.

We then use the \texttt{airball}\footnote{{airball.readthedocs.io} \citep{airball}} package implementation of the Chabrier 2005 \citep{chabrier_galactic_2003} initial mass function (IMF). We sample the IMF's probability distribution function to generate a distribution of stellar masses. These masses are then evolved to the present day using the MIST evolutionary tracks \citep{choi_mesa_2016} using the \texttt{minimint}\footnote{{github.com/segasai/minimint} \citep{koposov_segasaiminimint_2023}} package, assuming a metallicity of [Fe/H] = $-3.36$ dex and an age of 13.5 Gyr, which also returns the corresponding magnitudes for each star in the specified filter bandpasses. Finally, by scaling the total luminosity of this stellar population by the ratio of the number of stars we find and predict between $r \in [16, 20.5]$, that the luminosity of the stream in the $\phi_1 \in [-10^\circ, 40^\circ]$ segment is approximately $3\times10^3 \,\text{L}_\odot$. If we use the mass-to-light ratio of 2 assumed in \citet{martin_stellar_2022}, we find a stellar mass of $\sim6 \times 10^3 \text{ M}_\odot$. Because we only see a portion of the total stream, this can only be a lower bound on the mass of C-19's progenitor. Our measurement is in agreement with the estimated mass over a comparable segment of the stream in \citet{martin_stellar_2022}. Over the full known extent of the stream, \citet{yuan_pristine_2025} finds a lower-limit for the mass to be on the order of $10^4 \text{ M}_\odot$. 

\subsubsection{Heating Mechanisms}\label{sec:sims}

The velocity dispersion of C-19 is large compared to the typical GC stream, but our measurement of $7.8^{+1.5}_{-1.3} \text{ km s}^{-1}$ is cooler than the most recent measurements by \citet{yuan_pristine_2025}, which measures a velocity dispersion of $11.1^{+1.9}_{-1.6} \text{ km s}^{-1}$. We do note that the velocity dispersion varies along the extent of the stream, as we see in Figure \ref{fig:vdisp}, complicating a direct comparison between measurements. Regardless, due to the complex dynamical histories experienced by clusters, it is a challenge to make claims as to the origin of such heating. 

It has been proposed that C-19's progenitor was instead a dark-matter dominated DG \citep{errani_structure_2022,errani_pristine_2022}, which could account for the large dispersion, however the variations in light element abundances in C-19 member stars found in \citet{martin_stellar_2022} are indicative of a GC progenitor. Simulations of C-19 presented in \citet{Carlberg25_C19} used a single dominant halo containing CDM subhalos, finding that orbiting a GC progenitor for 12 Gyr produced streams with velocity dispersions comparable to our C-19 measurements. The large dispersion is also supported by the fact that the segment of C-19 we observe is currently nearing apocenter (see Figure \ref{fig:footprint}), where the stream is not as cool as it would be at pericenter. This occurs because near apocenter, the stream's orbital speed is lower, causing stars to `pile up' on-sky, mixing stars that span a broader range of orbital phases and energies. Alternatively, scenarios in which the progenitor was accreted alongside a dark-matter subhalo, possibly undergoing heating prior to its lifetime within the MW's halo, could assist in heating the stream \citep{carlberg_density_2020, malhan_new_2022}. Another possible explanation concerns the effects of disk tilting due to massive mergers (e.g. Gaia-Sausage-Enceladus) on stellar streams. Depending on the stream’s orbital inclination and the specifics of the disk’s reorientation in response to a merger, the stream’s width can be modified, becoming either narrower or broader \citep{dillamore_merger-induced_2022,nibauer_slant_2024}. 

Separately, an observational contribution to the large dispersion could arise from unresolved binaries, which can artificially inflate the inferred velocity dispersion. Of our member stars, 22 have multiple-epoch radial-velocity measurements; for one of these (* in Table \ref{tab:members}), the variability is significant enough to reject a constant-velocity model at the 99.9 \% level, with peak-to-peak variations suggesting possible binarity. Removing this star does not change our velocity-dispersion measurements along the stream, but a more complete accounting of binary contamination in our high-probability member sample would be required to assess the contribution of binaries to the observed dispersion.


\section{Conclusion}\label{sec:conclusion}
In this work we demonstrate our ability to characterize known stellar streams and identify new member stars based on metallicities and line-of-sight velocity measurements from three years of MWS observations, equivalent to the upcoming DESI DR2. We fit a mixture of four-dimensional truncated-Gaussian populations to the field about the metal-poor C-19 stellar stream. We then take the median of the posterior samples drawn by our MCMC sampler as the best-fit model parameters. We use a simple two-component model to characterize the stream and background populations in anticipation of applying this technique to many streams in DESI while needing minimal adjustments between streams.

From our fitted model, we assign membership probabilities to 2,071 stars, designating the 41 stars with probabilities $> 50\%$ as high-probable members. 35 of these 41 stars are new spectroscopically confirmed member stars, and represent a $\sim 2\times$ larger sample of spectroscopically confirmed stars than previous works. Moreover, we perform box-cuts about our spline tracks to identify horizontal branch member stars; we add 6 blue horizontal branch candidate member stars. 

Our large sample of member stars reveals substructure along C-19's length, including a sparse region between $\phi_1 \in (0^\circ, 20^\circ)$, and a spur between $\phi_1 \in (20^\circ, 35^\circ)$. We validate these structures are real by comparing the DESI MWS's completeness to \textit{Gaia} DR3 + DECaLS; the substructure appears in this deeper dataset, and the spur is observed in a relatively high-completeness area of the MWS's footprint.

 We can summarize our key findings as:

\begin{itemize}[label={-}]
    \item DESI is a powerful instrument to make robust measurements of stream properties and to identify new member stars within a stellar stream.
    \item We identify 41 high-probability member stars in the C-19 stellar stream along the MS+RGB isochrone, along with 6  members on the horizontal branch.

    \item We discover a spur associated with C-19, spatially offset from the main stream track.

    \item The spur members have a 10 km s$^{-1}$ offset from the main track, and a physical offset of $\sim 1^{\circ}$ corresponding to $\sim 300$ pc for a heliocentric distance of $18$ kpc. 
    
    \item Our stream model has velocity dispersion of $\sigma_{v_\text{GSR}}= 7.8_{-1.3}^{+1.5} \text{ km s}^{-1} $.
    
    \item Our stream model has a mean metallicity of $[\text{Fe/H}] = -3.36_{-0.10}^{+0.12} $ with a dispersion $\sigma_\text{[Fe/H]} <0.23_{-0.04}^{+0.05} \text{ dex}$.
    \item The luminosity of C-19 within $\phi_1 \in  (-10^\circ, 40^\circ)$ is $\approx 3\times10^{3} \text{ L}_\odot$, corresponding to a mass of $\sim 6\times10^3 \text{ M}_\odot$ at a mass-to-light ratio of 2.
\end{itemize}

C-19 is the first of many stellar streams we plan to analyze within the DESI MWS footprint. Measuring velocity dispersions across a population of streams can place robust constraints on cold versus warm DM models \citep{nibauer_textttstreamsculptor_2024, Carlberg25_C19}. In the future, we will extend this analysis to a larger sample of stellar streams, working toward stronger constraints on the nature of dark matter.

\begin{center}
    \textbf{Data Availability}
\end{center}

This work is supplemented by a Zenodo repository, from which the data behind the figures, all DESI stars assigned a membership probability greater than 0.1\%, BHB member stars, and MCMC posterior chains from our 27-parameter mixture model are accessible: \href{https://doi.org/10.5281/zenodo.20274724}{{10.5281/zenodo.20274724}}.

\begin{center}
    \textbf{Acknowledgements}
\end{center}
    NM thanks the LSST-DA Data Science Fellowship Program, which is funded by LSST-DA, the Brinson Foundation, the WoodNext Foundation, and the Research Corporation for Science Advancement Foundation; his participation in the program has benefited this work. NM was supported by the NSERC Canadian Graduate Masters Scholarship.

    N. M. and T.S.L. acknowledge financial support from Natural Sciences and Engineering Research Council of Canada (NSERC) through grant RGPIN-2022-04794.
    S.K. acknowledges support from Science \& Technology Facilities Council (STFC) (grant ST/Y001001/1).

    LBeS acknowledges support from CNPq (Brazil) through a research productivity fellowship, grant no. [304873/2025-0].
    
  This material is based upon work supported by the U.S. Department of Energy (DOE), Office of Science, Office of High-Energy Physics, under Contract No. DE–AC02–05CH11231, and by the National Energy Research Scientific Computing Center, a DOE Office of Science User Facility under the same contract. Additional support for DESI was provided by the U.S. National Science Foundation (NSF), Division of Astronomical Sciences under Contract No. AST-0950945 to the NSF’s National Optical-Infrared Astronomy Research Laboratory; the Science and Technology Facilities Council of the United Kingdom; the Gordon and Betty Moore Foundation; the Heising-Simons Foundation; the French Alternative Energies and Atomic Energy Commission (CEA); the National Council of Humanities, Science and Technology of Mexico (CONAHCYT); the Ministry of Science, Innovation and Universities of Spain (MICIU/AEI/10.13039/501100011033), and by the DESI Member Institutions: \url{https://www.desi.lbl.gov/collaborating-institutions}. Any opinions, findings, and conclusions or recommendations expressed in this material are those of the author(s) and do not necessarily reflect the views of the U. S. National Science Foundation, the U. S. Department of Energy, or any of the listed funding agencies.

The authors are honored to be permitted to conduct scientific research on I'oligam Du'ag (Kitt Peak), a mountain with particular significance to the Tohono O’odham Nation.

    This research has made use of the SIMBAD database, operated at CDS, Strasbourg, France \citep{wenger_simbad_2000}.
    This research has made use of NASA’s Astrophysics Data System Bibliographic Services.
    
    This paper made use of the Whole Sky Database (wsdb) created by Sergey Koposov and maintained at the Institute of Astronomy, Cambridge by Sergey Koposov, Vasily Belokurov and Wyn Evans with financial support from the Science \& Technology Facilities Council (STFC) and the European Research Council (ERC).
    
    This work has made use of data from the European Space Agency (ESA) mission
    {\it \textit{Gaia}} (\url{https://www.cosmos.esa.int/gaia}), processed by the {\it \textit{Gaia}}
    Data Processing and Analysis Consortium (DPAC,
    \url{https://www.cosmos.esa.int/web/gaia/dpac/consortium}). Funding for the DPAC
    has been provided by national institutions, in particular the institutions
    participating in the {\it \textit{Gaia}} Multilateral Agreement.

    For the purpose of open access, the author has applied a Creative
    Commons Attribution (CC BY) licence to any Author Accepted
    Manuscript version arising from this submission. 

    The DESI Legacy Imaging Surveys consist of three individual and complementary projects: the Dark Energy Camera Legacy Survey (DECaLS), the Beijing-Arizona Sky Survey (BASS), and the Mayall z-band Legacy Survey (MzLS). DECaLS, BASS and MzLS together include data obtained, respectively, at the Blanco telescope, Cerro Tololo Inter-American Observatory, NSF’s NOIRLab; the Bok telescope, Steward Observatory, University of Arizona; and the Mayall telescope, Kitt Peak National Observatory, NOIRLab. NOIRLab is operated by the Association of Universities for Research in Astronomy (AURA) under a cooperative agreement with the National Science Foundation. Pipeline processing and analyses of the data were supported by NOIRLab and the Lawrence Berkeley National Laboratory (LBNL). Legacy Surveys also uses data products from the Near-Earth Object Wide-field Infrared Survey Explorer (NEOWISE), a project of the Jet Propulsion Laboratory/California Institute of Technology, funded by the National Aeronautics and Space Administration. Legacy Surveys was supported by: the Director, Office of Science, Office of High Energy Physics of the U.S. Department of Energy; the National Energy Research Scientific Computing Center, a DOE Office of Science User Facility; the U.S. National Science Foundation, Division of Astronomical Sciences; the National Astronomical Observatories of China, the Chinese Academy of Sciences and the Chinese National Natural Science Foundation. LBNL is managed by the Regents of the University of California under contract to the U.S. Department of Energy. The complete acknowledgments can be found at https://www.legacysurvey.org/acknowledgment/.

\bibliography{bib}
\bibliographystyle{aasjournalv7}

\appendix

\section{Model Priors}\label{app:prior}

\begin{table}[h]
\centering
\caption{Prior ranges adopted for all fitted parameters used in \S \ref{sec:mcmc}.}
\begin{tabular*}{0.7\linewidth}{@{\extracolsep{\fill}} l c}

Parameter & Prior Range \\
\hline
\multicolumn{2}{l}{\small\textit{Stream component}} \\[0.5pt]

$f_s$ (\%)
  & $\mathcal{U}(10^{-4}, 1)$ \\

$E_{v_{\mathrm{GSR}},i}$ (km\,s$^{-1}$)  
  & $\mathcal{U}(-100, 100)$ \\

$E_{\mu_{\alpha,i}}$ (mas\,yr$^{-1}$)                 
  & $\mathcal{U}({-3, 3})$ \\

$E_{\mu_{\delta,i}}$ (mas\,yr$^{-1}$)                 
  & $\mathcal{U}({-8, 4})$ \\[4pt]

$\log\!\left(\sigma_{v_{\mathrm{GSR}}}/(\mathrm{km\,s^{-1}})\right)$

  & $\mathcal{U}(-2,\; 2)$ \\

$E_{\mathrm{[Fe/H]}}$ (dex)
  & $\mathcal{U}(-4,\; -2)$ \\

$\log \sigma_{\mathrm{[Fe/H]}}$ (dex)
  & $\mathcal{U}(-1,\; 0.5)$ \\[4pt]
\hline

\multicolumn{2}{l}{\small\textit{Background component}} \\[0.5pt]

$E_{v_{\mathrm{GSR}}}$ (km\,s$^{-1}$)
  & $\mathcal{U}(-100,\; 200)$ \\

$\log\!\left(\sigma_{v_{\mathrm{GSR}}}/(\mathrm{km\,s^{-1}})\right)$
  & $\mathcal{U}(-2,\; 2.5)$ \\[4pt]

$E_{\mathrm{[Fe/H]}}$ (dex)
  & $\mathcal{U}(-4,\; 1)$ \\

$\log\sigma_{\mathrm{[Fe/H]}}$ (dex)
  & $\mathcal{U}(-2,\; 0.5)$ \\[4pt]

$E_{\mu_\alpha}$ (mas\,yr$^{-1}$)
  & $\mathcal{U}(-4,\; 4)$ \\

$\log\!\left(\sigma_{\mu_\alpha}/(\mathrm{mas\,yr^{-1}})\right)$
  & $\mathcal{U}(-2,\; 1.5)$ \\[4pt]

$E_{\mu_\delta}$ (mas\,yr$^{-1}$)
  & $\mathcal{U}(-8,\; 0)$ \\

$\log\!\left(\sigma_{\mu_\delta}/(\mathrm{mas\,yr^{-1}})\right)$
  & $\mathcal{U}(-2,\; 1.5)$ \\

\hline
\end{tabular*}

\vspace{4pt}
\raggedright
\footnotesize
\label{tab:priors}
\end{table}

\section{Model Diagnostics}\label{app:ppc}

\begin{figure}[h]
    \centering
    \includegraphics[width=\linewidth]{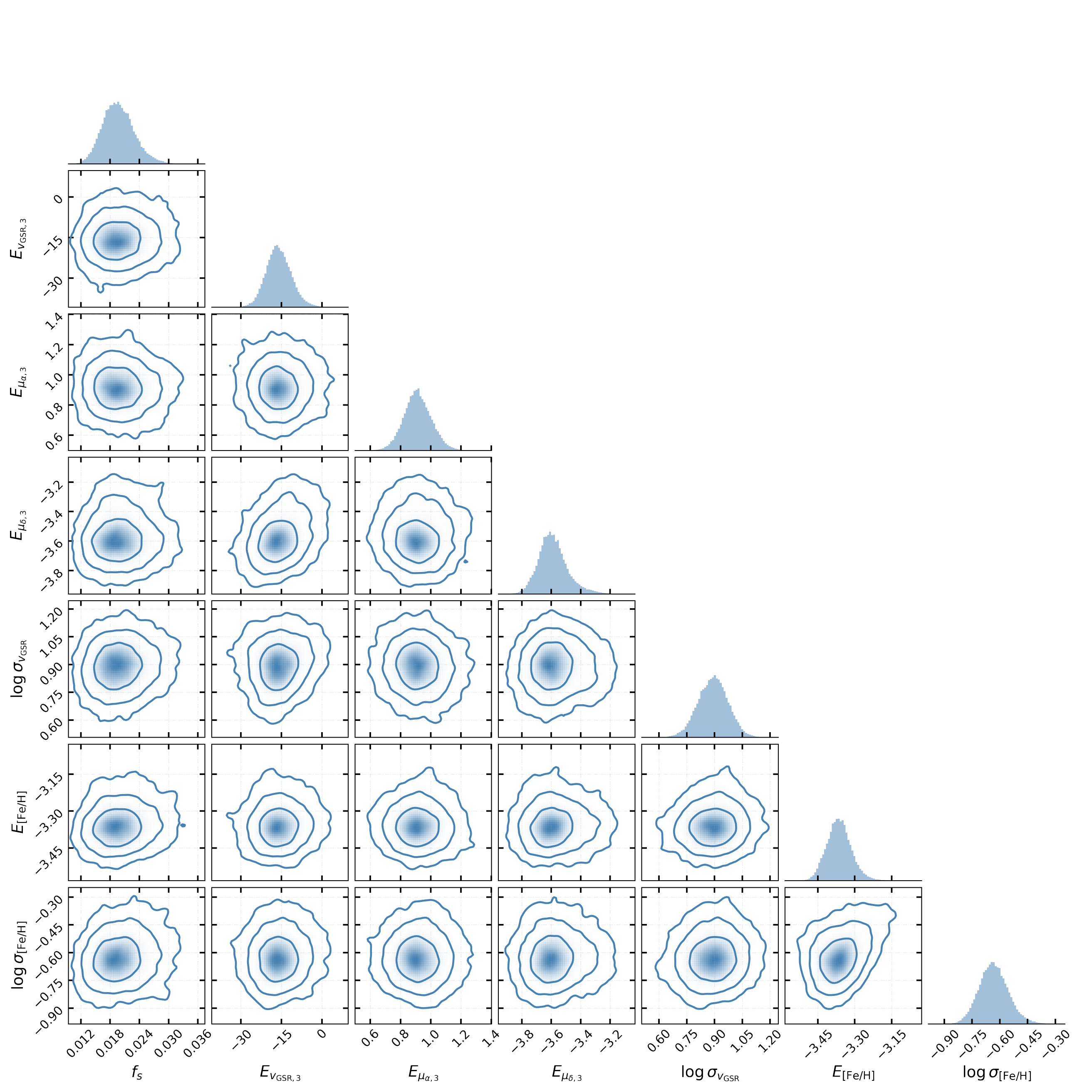}
    \caption{Subset of the 2D histogram corner plot of our posterior distributions, sampled using the \texttt{emcee} package \citep{foreman-mackey_emcee_2013}. Diagonal panels show the one-dimensional marginalized posteriors, and off-diagonal panels show the corresponding two-dimensional credible regions, illustrating parameter covariances and degeneracies. Our best-fit model parameters are chosen from the medians of these distributions.}
    \label{fig:corner_stream}
\end{figure}

\begin{figure}
    \centering
    \includegraphics[width=0.85\linewidth]{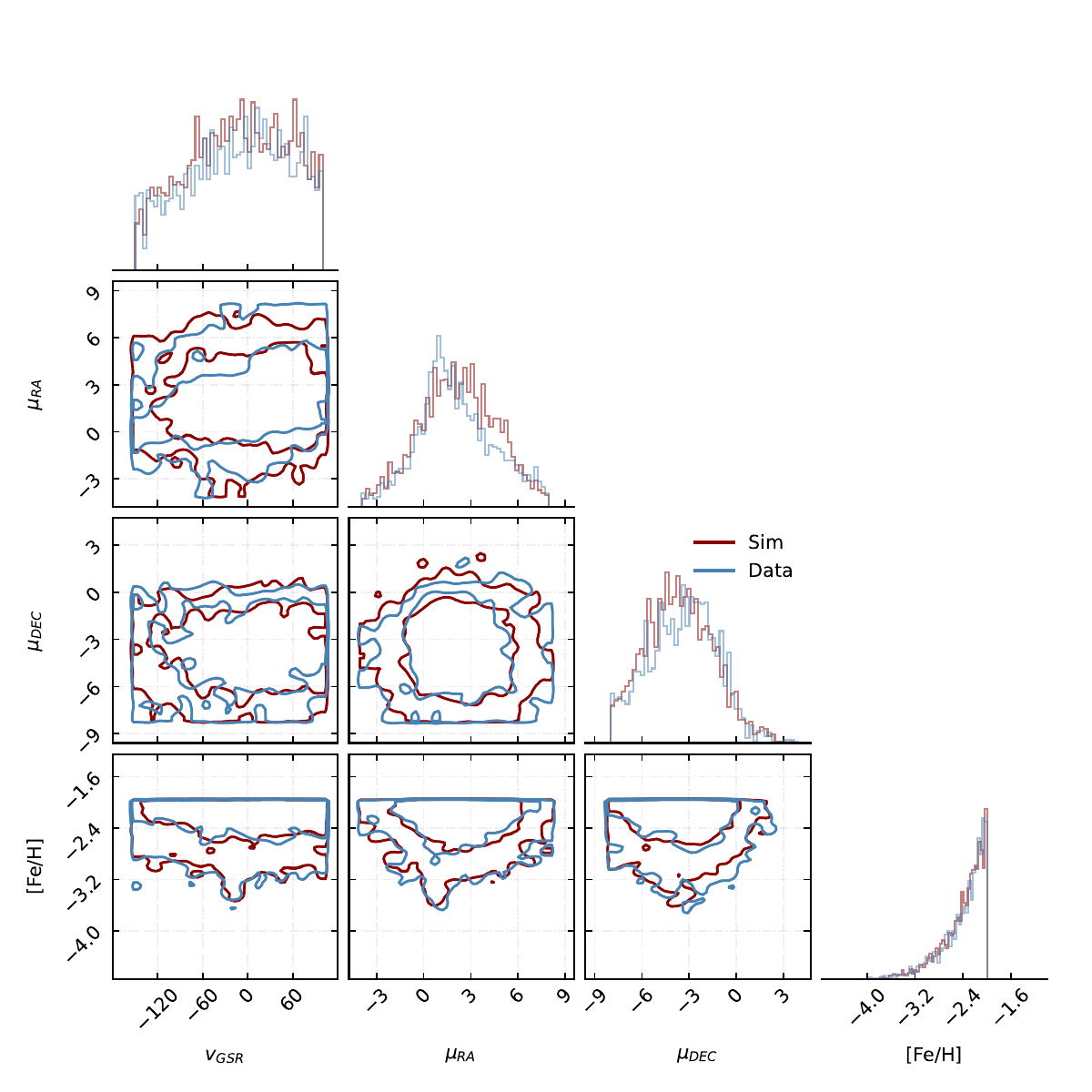}
    \caption{Corner plot comparing the observed catalogue \textit{(blue)} to mock data from a single posterior predictive realization \textit{(red)}. The mock catalogue is generated by drawing one sample from the MCMC posterior with added measurement uncertainties. The off-diagonal panels show the 2-dimensional density distribution of both the observed and simulated data. The diagonal panels show the one-dimensional distributions of each parameter in the real and simulated data. The overall agreement between the the two datasets indicates that a representative posterior draw reproduces the join distribution of the background and stream populations.}
    \label{fig:ppc_all}
\end{figure}

The posterior distributions of stream parameters from our analysis are shown as a corner plot in Figure \ref{fig:corner_stream}, which summarizes both the one-dimensional posteriors (diagonal panels) and the covariance between parameters (off-diagonal panels). To evaluate whether the model reproduces the observed background and stream populations, we generate a single posterior predictive realization, drawing from the posterior and the reported measurement uncertainties. This produces a mock catalogue that we compare to the observed data in Figure \ref{fig:ppc_all}. The figure overlays the two-dimensional distributions of the observed and mock population, with the one-dimensional distributions shown above. Taken together, this comparison show that a representative posterior draw generates a mock catalogue broadly consistent with the observations. This supports using the inferred posterior to draw conclusions about the stream (e.g., membership probabilities and intrinsic dispersions), since the generative model is not obviously mis-specified relative to the observed data. 

\begin{figure}
    \centering
    \includegraphics[width=0.95\linewidth]{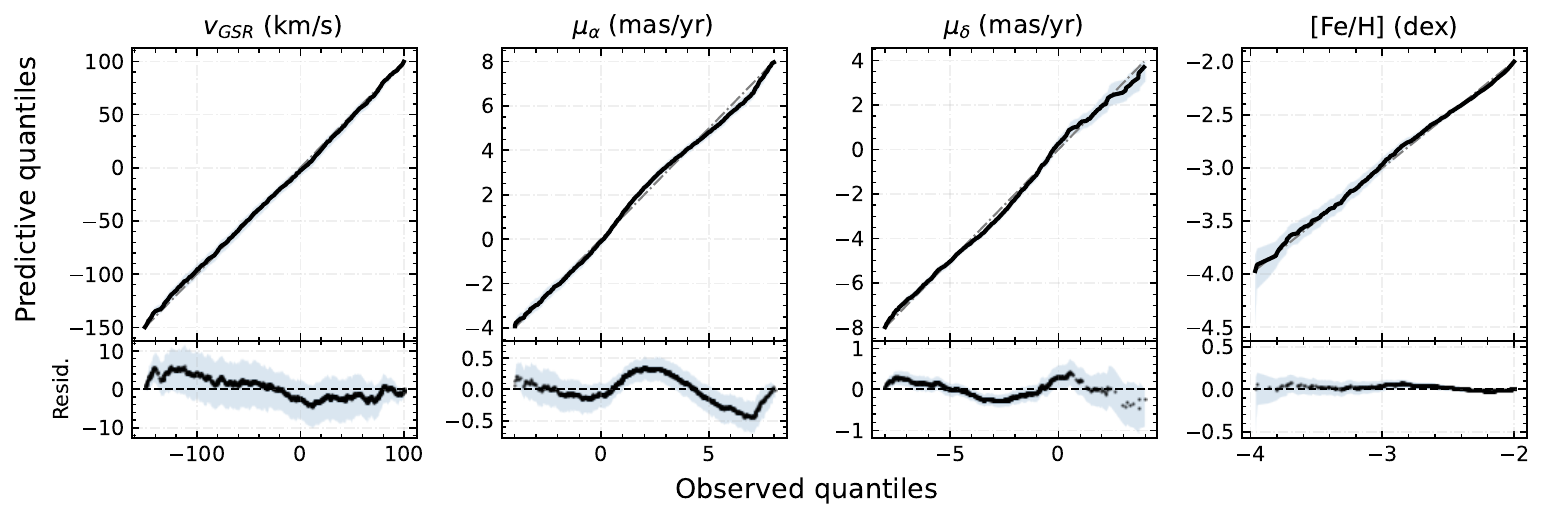}
    \caption{Quantile–quantile (Q–Q) plots comparing 500 posterior predictive (mock) quantiles to the observed quantiles for $v_{\mathrm{GSR}}$, $\mu_{\alpha}$, $\mu_{\delta}$, and $\mathrm{[Fe/H]}$. The mock catalogues are generated by drawing from the fitted model and propagating the reported observational uncertainties; the shaded band shows the 95\% posterior predictive interval and the solid line shows the 1:1 relation. The lower panels show residuals relative to the 1:1 line, where systematic departures would indicate a mismatch in distributional shape.}
    \label{fig:ppc_qq}
\end{figure}

We also show the quantile-quantile (Q-Q) plot for each of the four modelled parameters in Figure \ref{fig:ppc_qq}. Unlike the single illustrative realization in Figure \ref{fig:ppc_all}, we now use an ensemble of 500 posterior predictive draws, each generated by sampling from the posterior and perturbing by the reported measurement uncertainties. For each parameter, we sort the observed sample to obtain its empirical quantiles and, for each posterior predictive realization, sort the simulated sample and evaluate it on the same set of quantile levels as the data. The shaded regions represent the 95\% posterior predictive interval of the predicted quantiles after propagating observational uncertainties. The observed quantiles (black curve) lie close to the 1:1 line and within the predictive bands, indicating that the model’s predictive distribution is consistent with the data. Systematic departures from the 1:1 relation would signal a mismatch in distributional shape (e.g., skewness or heavier tails than the assumed Gaussian components).

\section{C-19 Member Stars}\label{app:mems}

\begin{figure}
    \centering
    \includegraphics[width=0.75\linewidth]{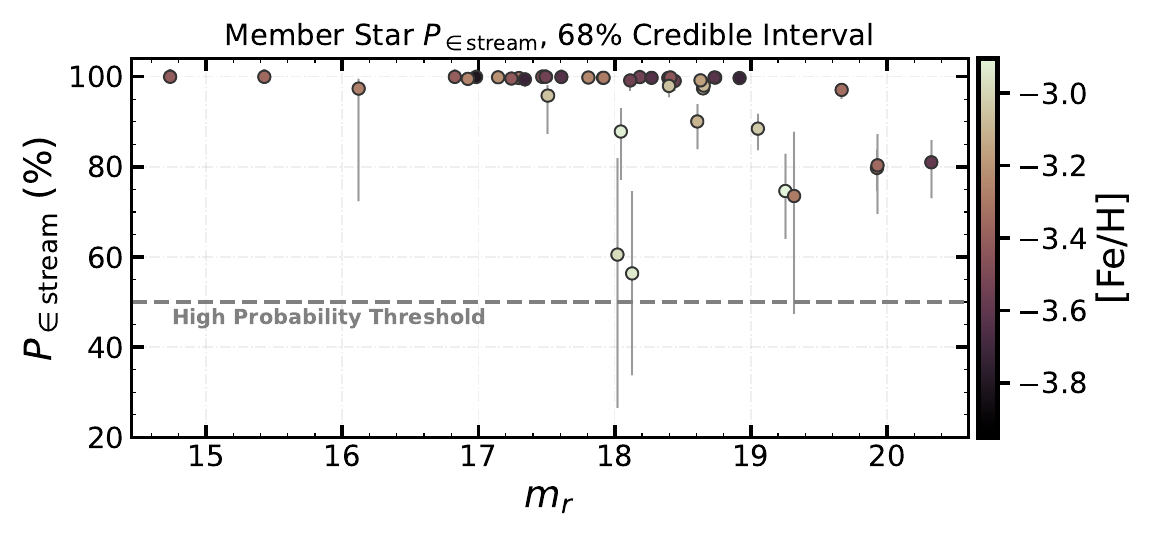}
    \caption{Member stars ($P_{\in \text{stream}} > 50\%)$ and their $P_{\in \text{stream}}$ values coloured by their metallicity. Extremely metal-poor members all exhibit high membership probabilities, while the (relatively) more metal rich members have a wider spread in membership probabilities. Error bars represent 16th and 84th percentiles of posterior distribution.}
    \label{fig:mem_dist}
\end{figure}

We provide a list of all DESI stars assigned a membership probability $P_{\in \text{stream}} > 0.1 \%$ by our mixture model on Zenodo (see Data Availability). A subset of the stars and data columns are shown in Table \ref{tab:members}. We denote the unresolved binary determined in \S \ref{sec:sims} with an asterisk. In Figure \ref{fig:mem_dist}, we show member stars ($P_{\in \text{stream}} > 50 \%$) and their membership probabilities as a function of their r-band magnitude. We derive uncertainties for the membership probabilities from the 16th and 84th percentiles of the posterior distribution and include them as error bars in Figure \ref{fig:mem_dist}. The member stars are coloured by their metallicity; the most metal-poor members exhibit high membership probabilities. The relatively metal-rich members have a wider spread in membership probabilities, and typically have larger uncertainties in their median $P_{\in \text{stream}}$. 

We also provide a list of the 6 BHB member stars identified in \S \ref{sec:box-cuts} on Zenodo (see Data Availability). A subset of the columns are given in Table \ref{tab:members2}, where the distances to the BHB members are from the DR2 version of the \citet{bystrom_exploring_2024} BHB catalogue.

\begin{figure}
    \centering
    \includegraphics[width=1\linewidth]{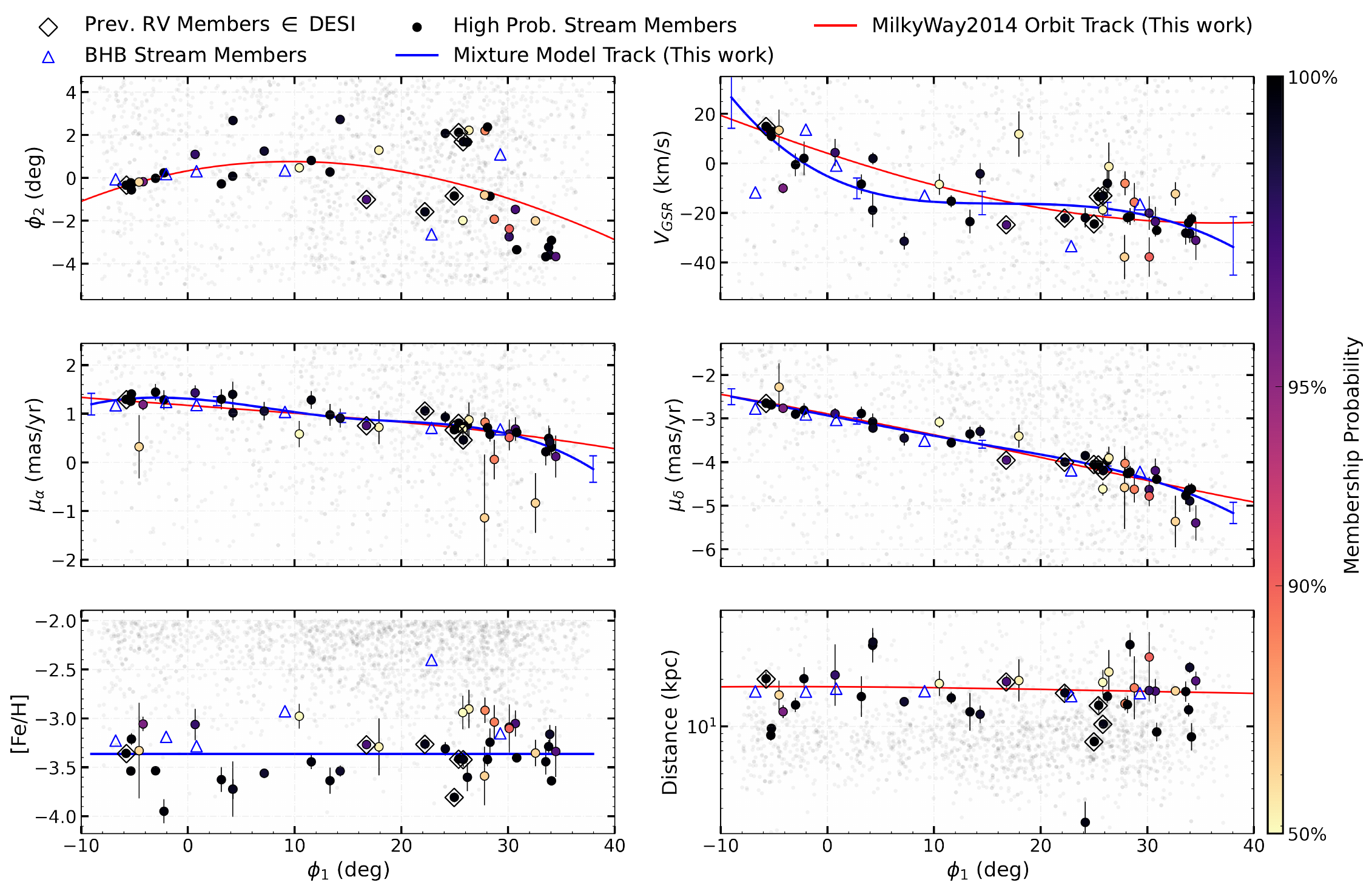}
    \caption{Identical to Figure \ref{fig:memprob}, except now shown in physical space rather than as residuals to the spline track.}
    \label{fig:6_app}
\end{figure}

In Figure \ref{fig:6_app} we recreate Figure \ref{fig:memprob}, except now we shown the spline tracks and member stars in physical space rather than as residuals to the spline track. 

\begin{table}
\centering
\caption{DESI stars with membership probabilities $P_{\in \text{stream}} > 0.01$ for C-19 members. Stars with membership probabilities greater than 0.5 are considered `high-probable' members in this work. The full table of 61 stars is accessible as a machine readable table or at \href{https://doi.org/10.5281/zenodo.20274724}{{doi.org/10.5281/zenodo.20274724}}.}
\label{tab:members}
\begin{tabular*}{\textwidth}{@{\extracolsep{\fill}} l l l c c c}
\hline
\textit{Gaia} Source ID & $\alpha$ (deg) & $\delta$ (deg) & [Fe/H] & $v_{\mathrm{{los}}}$ (km\,s$^{-1}$) & $P_{\in \text{stream}}$ \\
\hline
2644443254280033152 & 353.56322 & 0.28973 & $-3.31 \pm 0.07$ & $-149.7 \pm 3.9$ & 1.000 \\
$^*$2760807387346283648 & 351.31022 & 7.96340 & $-3.27 \pm 0.02$ & $-179.7 \pm 1.1$ & 0.970 \\
2641204161744171392 & 353.47032 & -0.97000 & $-3.42 \pm 0.03$ & $-137.4 \pm 1.3$ & 1.000 \\
2868052548930201984 & 354.77016 & 30.25098 & $-3.36 \pm 0.03$ & $-183.8 \pm 1.3$ & 1.000 \\
2812800131127935232 & 350.77770 & 12.93727 & $-3.02 \pm 0.09$ & $-191.3 \pm 3.6$ & 0.010 \\
2813219418719997824 & 353.52955 & 14.05124 & $-2.98 \pm 0.13$ & $-175.5 \pm 4.2$ & 0.610 \\
2632043095285833984 & 349.30774 & -5.83813 & $-3.05 \pm 0.13$ & $-139.0 \pm 6.0$ & 0.970 \\
2640793013114783872 & 352.99372 & -1.34662 & $-3.42 \pm 0.08$ & $-136.7 \pm 2.8$ & 1.000 \\
2640764352798121216 & 352.93297 & -1.72313 & $-3.60 \pm 0.14$ & $-130.6 \pm 5.7$ & 1.000 \\
2640695873838799232 & 353.46304 & -1.92645 & $-2.90 \pm 0.20$ & $-122.2 \pm 9.7$ & 0.750 \\
\hline
\multicolumn{6}{l}{\footnotesize $^{*}$Likely unresolved binary.}
\end{tabular*}
\end{table}


\begin{table}
\centering
\caption{BHB stars identified as members in \S \ref{sec:box-cuts} using stringent cuts about the mixture model's spline tracks. BHB stars are identified by cross-matching with the DR2 versions of the BHB catalogue \citep{bystrom_exploring_2024}. The table is accessible as a machine readable table or {at} \href{https://doi.org/10.5281/zenodo.20274724}{{doi.org/10.5281/zenodo.20274724}}.}
\label{tab:members2}
\begin{tabular*}{\textwidth}{@{\extracolsep{\fill}} l c c c c c}
\hline
\textit{Gaia} Source ID & $\alpha$ (deg) & $\delta$ (deg) & [Fe/H] & $v_{\mathrm{los}}$ (km\,s$^{-1}$) & $d$ (kpc) \\
\hline
2633283795373558272 & 352.02717 & -4.74327 & $-3.15 \pm 0.02$ & $-130.9 \pm 1.3$ & 16.35 \\
2657662987523302912 & 349.02498 & 2.07528 & $-2.40 \pm 0.00$ & $-174.0 \pm 2.2$ & 15.67 \\
2864750818590938624 & 354.82156 & 26.49102 & $-3.19 \pm 0.02$ & $-178.4 \pm 1.9$ & 16.76 \\
2871214881810792448 & 355.23572 & 31.21967 & $-3.23 \pm 0.01$ & $-208.2 \pm 1.5$ & 16.79 \\
2819555972950332416 & 353.56787 & 15.41058 & $-2.92 \pm 0.02$ & $-183.4 \pm 1.2$ & 16.86 \\
2828132236662858240 & 354.58016 & 23.64847 & $-3.28 \pm 0.13$ & $-187.8 \pm 1.8$ & 17.50 \\
\hline
\end{tabular*}
\end{table}

\end{document}